\documentclass[11pt,a4paper]{article}
\pdfoutput=1
\usepackage{jheppub}
\usepackage{amsmath,amssymb,dsfont}
\usepackage{verbatim}
\usepackage{graphicx}
\usepackage{mathrsfs}
\usepackage{appendix}
\usepackage{caption}
\usepackage{float}
\usepackage{enumerate}
\usepackage{subfig}
\usepackage{mathtools}
\usepackage{jheppub}
\usepackage{amsfonts}
\usepackage{ifthen}
\usepackage{amsthm}
\usepackage{tikz}
\usepackage{inputenc, array}
\usepackage{textcomp}
\usepackage[dvipsnames]{xcolor}

\usepackage{simplewick}
\usepackage{cancel}
\definecolor{mycolor}{RGB}{24, 116, 148}
\hypersetup{colorlinks=true,citecolor=mycolor,linkcolor=mycolor, linktocpage,urlcolor=mycolor}

\newcommand{\Z}{\mathcal{Z}}
\newcommand{\M}{\mathcal{M}}
\renewcommand{\Re}{{\rm Re\,}}
\renewcommand{\Im}{{\rm Im\,}}
\newcommand{\tr}[1]{{\,\rm tr}%
  \ifthenelse{\equal{#1}{}}{}{ \left( #1 \right) }%
}
\newcommand{\K}{\mathcal K}

\newcommand{\dv}[2]{\frac{d #1}{d#2}}

\usepackage{dynkin-diagrams}
\tikzset{/Dynkin diagram}

\newcommand{\trp}[1]{{\rm tr'}%
  \ifthenelse{\equal{#1}{}}{}{ \left( #1 \right) }%
}
\newcommand{\eval}[2]{\left. #1 \right|_{#2}}

\newcommand{\barz}{{\bar{z}}}
\renewcommand{\S}{\mathcal{S}}

\newcommand{\be}[1]{ \begin{equation}\label{#1} }
\newcommand{\ee}{\end{equation}}
\newcommand{\bea}[1]{\begin{eqnarray}\label{#1} }
\newcommand{\eea}{\end{eqnarray}}
\newcommand{\bes}{\begin{subequations}}
\newcommand{\ees}{\end{subequations}}

\newcommand{\D}{\mathcal{D}}

\DeclarePairedDelimiterX\braket[2]{\langle}{\rangle}{#1 \delimsize\vert #2}

\newcommand{\C}{\mathcal{C}}

\newcommand{\pdv}[2]{\frac{\partial#1}{\partial#2}}
\newcommand{\pdvsq}[2]{\frac{\partial^2#1}{\partial#2^2}}

\newcommand{\J}{\mathcal{J}}

\newcommand{\melonic}{\begin{tikzpicture}[baseline={(0,-0.1)},line width=0.2pt,scale=0.2]
  \draw (0,0) circle (1cm);
  \draw (-1,0) -- (1,0);
\end{tikzpicture}\,}

\renewcommand{\H}{\mathcal{H}}

\title{Saddling Timelike Liouville Theory}

\author{Emilie Despontin} 

\affiliation{Physique Théorique et Mathématique and International Solvay Institutes,\\
Université Libre de Bruxelles (ULB), C.P. 231, 1050 Brussels, Belgium.}
\emailAdd{\textcolor{mycolor}{emilie.despontin@ulb.be}}

\abstract{The two-sphere partition function of timelike Liouville theory admits two complementary semi-classical descriptions whose relation presents a non-perturbative puzzle. A direct path-integral expansion around the round two-sphere saddle reproduces the perturbative small-$\beta$ expansion, with $\beta$ the Liouville coupling, of the analytically continued Dorn--Otto--Zamolodchikov--Zamolodchikov (DOZZ) formula, while the latter contains an additional oscillatory factor suggestive of a second saddle contribution. We investigate the origin of this factor in the Liouville zero-mode sector using Picard--Lefschetz theory. Taking as input the Hankel-type contour motivated by the inverse-Gamma representation of the zero-mode integral, we determine its Lefschetz-thimble decomposition depending on $\beta$. For $0<\beta<1$, it precisely reproduces the saddle structure encoded in the analytically continued DOZZ result. We identify the integration contour as the source of the difference between the zero-mode saddle descriptions. As $\beta$ approaches $1$, the zero-mode saddle moves to the boundary of field space, while for $\beta>1$ the same contour is represented by a single contributing thimble. We further relate this Stokes reorganisation to recent probabilistic constructions of timelike Liouville theory.}

\keywords{Timelike Liouville theory, two-dimensional quantum gravity, Picard-Lefschetz theory, Stokes phenomena, Gaussian multiplicative chaos}

\begin{document}

\maketitle

\vfill

\newpage

\section{Introduction}

In the quest to understand four-dimensional Euclidean quantum gravity, one is tempted to go lower in dimensions. A promising laboratory has proven to be two-dimensional gravity coupled to conformal matter \cite{PolyakovNonCriticalStrings}. Upon fixing the metric to conformal gauge, the conformal factor is promoted to a dynamical field governed by Liouville theory \cite{Distler:1988jt,David:1988hj}. The resulting gauge-fixed theory consists of the matter, Liouville and ghost sectors, with the Liouville sector itself forming a two-dimensional conformal field theory (CFT). Vanishing of the conformal anomaly requires $c_m + c_L - 26 = 0$, where $c_m$ is the conformal matter central charge and $c_L = 1 + 6 \left(b+b^{-1}\right)^2$, with $b\in(0,1]$, the Liouville central charge. Accordingly, the Liouville coupling satisfies
\begin{equation}
    b = \frac{\sqrt{25-c_m}-\sqrt{1-c_m}}{2\sqrt{6}}.
\end{equation}

For $c_m\leq 1$, the Liouville coupling $b$ is real. This is the \textit{spacelike} Liouville CFT. It is a non-compact unitary CFT with a continuous spectrum of  primary states and non-normalisable vacuum. This framework has many applications in string theory \cite{PolyakovNonCriticalStrings,Moore:1991zv,Seiberg:2004at}, see also \cite{Collier:2023cyw,Collier:2024kmo} for recent results, and two-dimensional quantum gravity \cite{PolyakovNonCriticalStrings,Knizhnik:1988ak,Distler:1988jt,David:1988hj,Seiberg:1990eb,zamolodchikov_structure_1996}. One of the powerful aspects of Liouville theory is that its three-point structure constant $C_b$ is explicitly known, and called the \textit{Dorn--Otto--Zamolodchikov--Zamolodchikov (DOZZ) formula} \cite{Dorn_1994,zamolodchikov_structure_1996,Teschner:2001rv}. For $1<c_m<25$, the Liouville coupling becomes complex giving rise to the so-called \textit{complex} regime. For $c_m\geq 25$, $b$ becomes imaginary. This leads to \textit{timelike} Liouville theory, also called imaginary Liouville theory. \\

In recent years, interest in timelike Liouville theory has become increasingly prominent in the literature, in string theory as well as in cosmology \cite{Harlow_2011,Giribet:2011zx,Ribault:2015sxa,Mertens:2020hbs,anninos_two-sphere_2021,Collier:2023cyw,anninos_remarks_2025,Usciati:2025cdn,Anninos_2026,chatterjee2026rigorousresultstimelikeliouville,chatterjee2026exactcalculationschargeneutrality,Giribet:2026gao,Cercle:2026ihl,CarneirodaCunha:2003mxy,Bautista:2019jau}. Along these lines, the semi-classical expansion of the timelike Liouville sphere partition function about the round two-sphere saddle has been studied in \cite{anninos_two-sphere_2021,muhlmann_two-sphere_2021}, up to three-loop order. A finite-$\beta$ expression for the sphere zero-point function was also obtained using an analytically continued Coulomb-gas construction in \cite{Giribet:2022cvw}.  One should distinguish two appearances of timelike Liouville theory: timelike Liouville \textit{CFT} and timelike Liouville \textit{gravity}. On the gravitational side, timelike Liouville arises as the conformal mode of two-dimensional gravity after gauge fixing, with its parameters constrained by anomaly cancellation with the matter and ghost sectors, as mentioned above. It results in a real Euclidean action with a ``wrong-sign" kinetic term, leading to the \textit{timelike} name. This provides a two-dimensional analogue of the conformal mode problem of Euclidean gravity \cite{Gibbons:1978ac,polchinski_phase_1989,Maldacena:2024spf}. 
Timelike Liouville should be viewed as a distinct quantum theory rather than simply as spacelike Liouville evaluated at imaginary $b$.\\

On the CFT side, one would like to determine the spectrum and structure constants characterising the non-compact, non-unitary timelike Liouville CFT \cite{Zamolodchikov:2005fy,Kostov:2005av,Ribault:2015sxa}. The distinction with spacelike Liouville is already visible at the level of the spectrum. In spacelike Liouville theory, the physical continuum and the degenerate Virasoro representations occupy distinct loci in the complex momentum plane, whereas after continuing to the timelike regime, $b\to-i\beta$, the degenerate momenta are rotated onto the same locus as the physical spectrum. This reorganisation motivates treating the timelike conformal bootstrap as a distinct CFT problem, rather than simply evaluating the spacelike theory at imaginary $b$. Indeed, the generic spacelike DOZZ structure constant $C_b(\alpha_1,\alpha_2,\alpha_3)$, where the $\alpha_i$ denote the Liouville momenta, cannot in general be analytically continued to purely imaginary $b$. The obstruction can be traced to the analytic properties of the $\Upsilon_b$ function entering the DOZZ formula: its continuation becomes singular as $b$ approaches the imaginary axis, preventing a well-defined continuation of the generic structure constant \cite{Zamolodchikov:2005fy}. Nevertheless, the shift equations underlying the conformal bootstrap can themselves be extended through a Virasoro Wick-rotation \cite{Ribault:2023vqs} and admit a second solution, usually referred to as the timelike DOZZ structure constant $\widehat C_\beta$ \cite{Harlow_2011,Giribet:2011zx,Kostov:2005av,Zamolodchikov:2005fy,Ribault:2015sxa}.  However, when evaluated for three insertions of $\beta$, it vanishes, $\widehat C_\beta(\beta,\beta,\beta)=0$ \cite{anninos_two-sphere_2021,muhlmann_two-sphere_2021}. By contrast, the special spacelike structure constant $C_b(b,b,b)$ evades this obstruction to analytic continuation. For these particular values, the arguments of the $\Upsilon_b$ functions are related by their shift identities, allowing the latter to be eliminated in favour of ordinary $\Gamma$-functions before the continuation is performed. The resulting expression admits a non-trivial continuation under $b\to-i\beta$, which we denote by $\C(\beta,\beta,\beta)$. Remarkably, despite the distinction between the spacelike and timelike CFTs, this special analytic continuation of the spacelike DOZZ structure constant still appears to retain non-trivial information about the timelike theory, as suggested in \cite{anninos_two-sphere_2021,muhlmann_two-sphere_2022}.\\

A natural next step is to investigate to what extent these CFT data can be reproduced from a semi-classical path-integral perspective and in particular, for the purposes of this work, from the two-sphere partition function. How does the small-$\beta$ expansion of the analytically continued structure constant $\C$ emerge from the timelike path integral? The two descriptions can be naturally related through \cite{zamolodchikov_structure_1996,anninos_two-sphere_2021}
\begin{equation}
-\partial_\Lambda^3 \Z_{\rm tL}[\beta,\Lambda]
=
2\C(\beta,\beta,\beta).
\end{equation}
From the path-integral perspective, the difference between the spacelike and timelike constructions is expected to be encoded, at least in part, in the choice of integration cycle in the timelike Liouville field space. Previous perturbative calculations reveal a striking relation between the two descriptions: the direct timelike sphere path integral and the analytically continued DOZZ expression \cite{anninos_two-sphere_2021,muhlmann_two-sphere_2022} reproduce the same perturbative expansion about the round two-sphere saddle. However, the latter approach contains an additional oscillatory factor, naturally suggestive of an additional saddle contribution. This raises the question of whether this contribution can be understood directly from the saddle decomposition of an appropriate timelike integration cycle.\\

Some steps toward understanding this mismatch have been taken in both the physics and mathematics communities \cite{Lacoin:2013joa,guillarmou2023compactifiedimaginaryliouvilletheory,Cao_2023,Usciati:2025cdn,chatterjee2026rigorousresultstimelikeliouville,chatterjee2026exactcalculationschargeneutrality,Cercle:2026ihl}. In this paper, motivated by analytic continuation and by the inverse-Gamma representation \cite{Harlow_2011,Cao_2023}, we take a Hankel-type contour for the Liouville zero mode as our starting point and use Picard--Lefschetz theory to determine its convergent steepest-descent cycle, or \textit{Lefschetz thimble}, decomposition. The inverse-Gamma integration cycle and its two-thimble factor were analysed in Appendix~C of \cite{Harlow_2011}. Here we apply that mechanism to the timelike sphere zero-point integral. For the chosen zero-mode cycle, we compute the dual-thimble intersection numbers on both sides of $\beta=1$ and compare the
resulting saddle sectors with the proposed sphere partition function and the round-sphere loop expansion. In the regime $0<\beta<1$, with $b=-i\beta$, we find that this contour receives contributions from two thimbles. Their relative weighting reproduces the saddle structure suggested by the oscillatory factor appearing in the analytically continued DOZZ result. Our Picard--Lefschetz analysis is restricted to the zero mode and therefore determines the saddle sectors and their relative phases, while the non-zero modes account for the perturbative fluctuations around them.
%Using the Picard--Lefschetz theory, we investigate how the choice of integration contour for the sphere partition function determines the contributing saddle sectors, and whether an appropriate contour can account for the missing saddle contribution. We find that, once the zero-mode integral is placed on the Hankel-type contour suggested by analytic continuation and by the inverse-Gamma representation \cite{Harlow_2011,Cao_2023}, its Picard--Lefschetz decomposition contains two thimbles in the $0<\beta<1$ regime, where $b=-i\beta$. Our Picard--Lefschetz analysis is restricted to the zero mode: it identifies the saddle content controlling the semi-classical exponential and its relative phases, while the non-zero modes supply the usual perturbative determinants and loop corrections. 
For $0<\beta<1$, the real zero-mode contour converges, so choosing the Hankel-type contour is an additional prescription rather than a consequence of convergence. For $\beta>1$, the real zero-mode integral diverges and requires an additional definition, but this does not uniquely determine a complex integration cycle. We stress that Picard--Lefschetz theory determines the thimble decomposition once an integration cycle has been specified. Related contour prescriptions appear in probabilistic constructions of timelike Liouville theory. They provide an independent and complementary non-perturbative perspective on the definition of the theory \cite{Usciati:2025cdn,chatterjee2026exactcalculationschargeneutrality,Cercle:2026ihl}. They motivate a comparison, although they do not by themselves fix the sphere zero-point contour. Establishing a precise dictionary between these constructions and the Picard--Lefschetz description would help clarify which saddle decomposition corresponds to the physically relevant path integral. \\

The paper is organised as follows. In Sec. \ref{Sec:TimelikeLiouville}, we review the two-sphere timelike Liouville partition function and the comparison between the direct path-integral result and the analytically continued DOZZ expression.  In Sec. \ref{Sec:MorseTheory}, we recall the Picard--Lefschetz ingredients needed for the analysis. In Sec. \ref{Sec:SaddleContribtL}, we apply these tools to the timelike Liouville zero mode, determine the thimble structure in the regimes $0<\beta<1$ and $\beta>1$, and discuss the degenerate point $\beta=1$.  Finally, in Sec. \ref{Sec:GMC}, we relate the contour prescription to recent probabilistic constructions of imaginary Liouville theory and distinguish the Picard--Lefschetz saddle question from the imaginary Gaussian multiplicative chaos (iGMC) and charge-neutrality bounds.\\

\section{Timelike Liouville and the Two-Sphere Partition Function}
\label{Sec:TimelikeLiouville}
Timelike Liouville theory arises in two-dimensional gravity coupled to a conformal matter CFT with a central charge $c_m >25$ \cite{Polchinski:1989fn}. 
The relation to the spacelike parametrisation follows from anomaly cancellation. The total worldsheet central charge obeys
\begin{equation}
    c_m + c_L - 26 = 0, \qquad c_L = 1+6 Q^2,
\end{equation}
leading to
\begin{equation}
    Q^2 = \frac{25-c_m}{6}.
\end{equation}
Hence, starting from the spacelike parametrisation $Q=b+b^{-1}$ and going into the complex-value regime
\begin{equation}
    q = - i Q, \qquad \beta = i b,
\end{equation}
one obtains on the branch admitting the semi-classical limit $\beta\to0$
\begin{equation}
    Q = \sqrt{\frac{25-c_m}{6}}, \qquad b=\frac{\sqrt{25-c_m}-\sqrt{1-c_m}}{2\sqrt 6}.
\end{equation}
In other words, $Q$ and $b$ become purely imaginary while $q$ and $\beta$ are real. The gravitational path integral takes the schematic form
\begin{equation}
    \Z_{\rm grav.}[\Lambda] = \sum_{h=0}^\infty e^{\vartheta \chi_h} \int \D g_{ij} \, e^{-\Lambda \int_{\Sigma_h}d^2x \, \sqrt{g}} \Z_{\rm CFT}^{(h)}[g_{ij}],
\end{equation}
where $\chi_h$ is the Euler characteristic of genus-$h$ surface $\Sigma_h$, $\vartheta$ is the bare gravitational coupling and $\Z_{\rm CFT}^{(h)}$ the matter CFT partition function. The condition $\Lambda>0$ ensures that configurations with large total area are exponentially suppressed in the Euclidean path integral.\\ %{\color{red} read the beginning of my paper with teresa and dio. It is explained tere what happens to the CFT. } 

We restrict throughout to the sphere topology, corresponding to the leading contribution in the genus expansion. Writing
\begin{equation}
g_{ij}=e^{2\beta\phi}\tilde g_{ij},
\end{equation}
and integrating out the matter and the Faddeev--Popov ghosts, their Weyl anomalies are incorporated into the Liouville action, while their remaining dependence on the fiducial metric contributes only an overall $\phi$-independent factor. Up to this  normalisation and the genus-zero topological factor, the gravitational path integral therefore reduces to
\begin{equation}
\Z_{\rm grav.}^{(0)}
\sim
\Z_{\rm tL}
=
\frac{1}{{\rm vol}_{PSL(2,\mathbb C)}}
 \int \D\phi\, e^{-\S_{\rm tL}[\phi]}.
 \label{Eq:tLPI}
\end{equation}
with the Euclidean timelike Liouville action 
\begin{equation}
    \S_{\rm tL}[\phi] = \frac{1}{4\pi}\int d^2x\, \sqrt{\tilde g}\left(-{\tilde g}^{ij}\partial_i\phi \partial_j\phi - q \tilde R \phi + 4\pi \Lambda e^{2\beta\phi}\right),
    \label{Eq:tLAction}
\end{equation}
where $\tilde g$ is the fiducial metric,  $\tilde R$ its Ricci scalar, $\Lambda>0$ is the cosmological constant, and
\begin{equation}
    q=\frac 1\beta- \beta, \quad \beta \in \mathbb R.
\end{equation}
The negative kinetic term in Eq. \eqref{Eq:tLAction} is the two-dimensional analogue of the conformal-mode problem in Euclidean gravity, making timelike Liouville theory a useful setting in which the contour and saddle structure of a gravitational path integral can be investigated explicitly, while retaining access to both semi-classical and conformal bootstrap techniques. The sphere partition function in Eq. \eqref{Eq:tLPI} has been evaluated perturbatively up to three-loop order \cite{anninos_two-sphere_2021,muhlmann_two-sphere_2021} by expanding around the constant saddle
\begin{equation}
\phi_*
=
\frac{1}{2\beta}
\log\left(
\frac{q}{4\pi v\Lambda\beta}
\right),
\end{equation}
where $4\pi v$ is the area of the fiducial two-sphere. In the semi-classical limit $\beta\to0$, it results in
\begin{align}
    \Z_{\rm tL}[\Lambda]&\approx \pm i e^{-\tfrac{1}{\beta^2}-\tfrac{1}{\beta^2}\log(4\pi\beta^2)}v^{\tfrac{c_{\rm tL}}{6}}\Lambda_{\rm uv}^{\frac76-\beta^2}\Lambda^{-\tfrac{1}{\beta^2}+1}\Bigg[\frac{1}{\beta}+\left(\frac{1}{6}(19-6 \log 4)-\left(2 \gamma_E+\log \pi\right)\right. \nonumber\\
    &\left.+\,\melonic-\frac{10}{3}+\frac{5 \pi a_1}{2 a_0}\right) \beta+\left(\frac{1}{2}\left(\frac{1}{6}(19-6 \log 4)-\left(2 \gamma_E+\log \pi\right)+\,\melonic-\frac{10}{3}+\frac{5 \pi a_1}{2 a_0}\right)^{\hspace{-3pt}2}\right.\nonumber \\
    &\left.+\operatorname{loops}_{\beta^4}-\frac{17}{27}+\frac{15 \pi a_1}{4 a_0}-\frac{25 \pi^2 a_1^2}{8 a_0^2}-\frac{1}{2} \,\melonic^2\right) \beta^3+\ldots\Bigg]
\end{align}
where $\gamma_E$ is the Euler--Mascheroni constant, $\Lambda_{\mathrm{uv}}$ is the UV cutoff of the theory, and $a_0, a_1$ encode regulator-dependent data. Choosing $a_1=27 a_0 /(20 \pi)$ implements the matching conventions used below. Finally, the symbol $\,\melonic$ denotes a \textit{melonic} two-loop diagram, while $\operatorname{loops}_{\beta^4}$ denotes the collection of all three-loop diagrams contributing at order $\mathcal{O}(\beta^4)$.\\

A complementary prediction for the sphere partition function follows from the conformal bootstrap. As mentioned in the introduction, although the generic spacelike DOZZ structure constant does not admit a straightforward continuation to purely imaginary $b$, the special quantity $C_b (b,b,b)$ admits a non-trivial continuation. Starting form the DOZZ formula of Dorn--Otto \cite{Dorn_1994} and Zamolodchikov--Zamolodchikov \cite{zamolodchikov_structure_1996}, one obtains under $b\to \pm i \beta$ \cite{Harlow_2011,Giribet:2011zx,anninos_two-sphere_2021}
\begin{align}
\mathcal{Z}^{\mathrm{DOZZ}}_{\rm tL}[\Lambda]
&\approx
\pm
e^{-\frac{1}{\beta^2}
-\frac{1}{\beta^2}\log(4\pi\beta^2)}
\Lambda^{-\frac{1}{\beta^2}+1}
\left(
1-e^{\frac{2 i \pi}{\beta^2}}
\right)\times \nonumber\\
& \times
\left[
\frac{1}{\beta}
+
\frac{1}{6}(19-6\log 4)\beta
+
\frac{1}{2}\left(
 \frac{1}{36}(19-6\log 4)^2
-
\frac{2}{3}\zeta(3)
\right)\beta^3
+\cdots
\right]
\label{Eq:tLDOZZ}
\end{align}
Moreover, Giribet and Leoni obtained a finite-$\beta$ expression for the timelike sphere zero-point function using an analytically continued Coulomb-gas computation \cite{Giribet:2022cvw}. The unexpanded expression
obtained by continuing the DOZZ formula in
\cite{anninos_two-sphere_2021}, whose small-$\beta$ expansion is displayed
in Eq.~\eqref{Eq:tLDOZZ}, has the same $\Lambda$ dependence. For compatible branch choices, the two expressions differ at fixed
$\Lambda$ by a phase and the $\beta$-dependent factor
$e^{(1-2\log 2)q^2}$. \\

The direct path-integral computation and the analytically continued DOZZ expression exhibit the same semi-classical structure: they share the leading exponential, the exact dependence $\Lambda^{-1/\beta^2+1}$, and a perturbative expansion in odd powers of $\beta$. A coefficient-by-coefficient comparison nevertheless requires distinguishing quantities of different origin. The factor $v^{c_{\rm tL}/6}$ is fixed by the Weyl anomaly of timelike Liouville theory and is cancelled by the corresponding matter and ghost contributions in the full gravitational partition function. Meanwhile, the coefficients $a_0$ and $a_1$ arise from the residual $PSL(2,\mathbb C)$ Faddeev--Popov determinant. By contrast, the overall normalisation and the finite definition of the ultraviolet scale are scheme-dependent. Upon identifying the schemes through the finite rescaling
\begin{equation}
\Lambda_{\rm uv}\longrightarrow
e^{-(2\gamma_E+\log\pi)}\Lambda_{\rm uv},
\end{equation}
the two-loop coefficient agrees with the analytically continued DOZZ prediction. At order $\beta^3$, the matching further relies on the diagrammatic identity proposed and numerically tested in \cite{muhlmann_two-sphere_2021}. The perturbative path-integral result is therefore consistent with the DOZZ expansion through the presently computed orders. The two- and three-loop calculations provide independent path-integral checks of the perturbative fluctuations about the round two-sphere saddle. They do not determine the full finite-$\beta$ partition function or the relative weights of other saddle sectors. Comparing them with the closed expressions isolates the part associated with this single saddle, leaving the additional factor
\begin{equation}
1-e^{2\pi i/\beta^2},
\label{Eq:TwoSaddleFactor}
\end{equation}
which is present in the analytically continued DOZZ result. Its form strongly suggests that the analytically continued result contains contributions from more than one saddle. Indeed, the two terms in Eq. \eqref{Eq:TwoSaddleFactor} differ by an exponentially weighted phase and hence naturally invite an interpretation in terms of distinct saddle sectors. The central question of this work is whether such an additional contribution follows from the integration cycle of the timelike Liouville path integral itself. We address this question by isolating the Liouville zero mode and determining the Leftschetz-thimble decomposition of an appropriate complex integration contour.\\

This contour question also appears in recent probabilistic constructions of imaginary, or
timelike, Liouville theory. Usciati, Guillarmou, Rhodes and Santachiara proposed a non-compactified imaginary Liouville theory based on a real Gaussian free field and argued that
it reproduces imaginary DOZZ structure constants without imposing a neutrality constraint
\cite{Usciati:2025cdn}. Chatterjee instead developed a rigorous formulation based on Gaussian variables of negative variance and derived timelike DOZZ-type formulas subject to charge-neutrality conditions \cite{chatterjee2026rigorousresultstimelikeliouville}. Subsequent work extended the analysis beyond charge neutrality in special regimes \cite{chatterjee2026exactcalculationschargeneutrality}. These constructions
provide complementary non-perturbative prescriptions for timelike Liouville theory. We return to them later and compare their conditions with the contour and saddle structure obtained from the zero-mode analysis.

\section{Morse Theory}
\label{Sec:MorseTheory}
In this section, we review the Morse-theoretic construction underlying Picard--Lefschetz theory, following \cite{Pasquetti_2010,witten2010analyticcontinuationchernsimonstheory}. We will recall how oscillatory integrals may be decomposed into canonical convergent integration cycles, called \textit{Lefschetz thimbles}, and how this decomposition changes when varying parameters. The typical object of interest is
\begin{equation}
    Z(\lambda) = \int_{\C (\lambda)} e^{-S(z,\lambda)}\,dz,
    \label{Eq:OscillatoryIntegral}
\end{equation}
where $S$ is complex-valued, usually holomorphic in $z$, and $\lambda$ denotes a set of external parameters. The cycle $\C$ must be chosen so that the integral is convergent. When the integrand is not absolutely convergent on the original contour, one analytically continues the problem by deforming the integration cycle into a region where the real part of the exponent decreases sufficiently fast at infinity. As parameters $\lambda$ vary, a contour that is admissible in one region of parameter space may fail to be admissible in another. One is therefore led to study integration cycles not as fixed contours, but as elements of suitable relative homology groups. In this language, the integral is unchanged under deformations of the cycle inside the same relative homology class. However, the decomposition of such a cycle into a preferred basis of convergent cycles can jump across special loci in parameter space. These jumps constitute the \textit{Stokes phenomenon} described by Picard--Lefschetz theory. \\

The basic idea is the following. Let $\mathcal P$ denote the parameter space. For each
$\lambda\in\mathcal P$, consider the holomorphic function
\begin{equation}
    S_\lambda:\mathcal M\to\mathbb C,
    \qquad
    S_\lambda(z)\equiv S(z,\lambda).
\end{equation}
In the following, $\lambda$ is initially held fixed and its dependence is suppressed. It will be restored when discussing Stokes phenomena. Assume, for simplicity, that the critical points of $S_\lambda$ are isolated and non-degenerate. We denote $\Sigma=\{p_\sigma \}_\sigma$ the set of critical points. To each critical point, one associates an integration cycle $\J_\sigma$, called a \textit{Lefschetz thimble}. These thimbles are constructed as steepest-descent cycles for the real part of the exponent. Their key property is that the integral
\begin{equation}
    \int_{\J_\sigma} dz\, e^{-S_\lambda}
\end{equation}
is convergent, provided that the thimble ends at infinity in regions where $\Re S_\lambda\to \infty$. Then, any admissible integration cycle $\C$ can be expanded in the thimble basis as
\begin{equation}
    \C = \sum_\sigma n_\sigma \J_\sigma,\qquad n_\sigma \in \mathbb Z.
\end{equation}
The coefficients $n_\sigma$ are determined by the dual upward thimbles $\K_\sigma$
\begin{equation}
    n_\sigma = \langle \C,\K_\sigma \rangle.
\end{equation}
where $\langle\cdot,\cdot\rangle$ denotes the oriented intersection pairing between admissible integration cycles and upward thimbles. Its relative-homology formulation will be given below. Thus, Picard--Lefschetz theory reduces the analytic problem of choosing convergent contours to a topological problem. One computes the relative homology class of the chosen integration cycle and expands it in a basis of thimbles.\\

\paragraph{Holomorphic Morse Theory.}
Let $\M$ be a complex manifold of complex dimension $n$, and let
\begin{equation}
    S_\lambda : \M \to \mathbb C
\end{equation}
holomorphic. Regarding $\mathcal M$ as a real manifold of dimension $2n$, we introduce the Morse function
\begin{equation}
    h = \Re S_\lambda.
\end{equation}
The gradient flow of $h$ defines steepest-descent trajectories for the integrand $e^{-S_\lambda}$. Given local holomorphic coordinates $z^i$, a critical point of $S_\lambda$ is a point $p$ satisfying 
\begin{equation}
    \eval{\pdv{S_\lambda}{z^i}}{p} = 0, \qquad \forall\, i=1, \dots, \dim \M.
\end{equation}
Because $S_\lambda$ is holomorphic, its critical points coincide with those of $h$. Locally near a non-degenerate critical point $p$, the holomorphic Morse lemma implies that one can choose local holomorphic coordinates $z^1,\dots,z^n$, with $n=\dim_{\mathbb C} \M$, such that
\begin{equation}
    S_\lambda = S_\lambda(p) + \frac 12 \sum_{i=1}^{n}(z^i)^2.
\end{equation}
This reflects the fact that, in the holomorphic story, positive and negative directions are paired within each complex coordinate. Now, write $z^i=x^i+i y^i$ such that
\begin{equation}
    h = h(p) + \frac12 \sum_{i=1}^{n} (x^i)^2 -\frac12 \sum_{i=1}^{n} (y^i)^2,
\end{equation}
giving $n$ descending and $n$ ascending directions associated with each critical point, and the Morse index is $n$. Let us write the downward flow equation explicitly. On the complex manifold $\M$, choose a Kähler metric $ds^2=g_{i{\bar j}}dz^i d\bar z^{\bar j}$. The downward gradient flow equations of $h=\Re S_\lambda$ are
\begin{equation}
    \frac{dz^i}{dt}
    =
    -g^{i\bar j}
    \frac{\partial \overline{S_\lambda}}{\partial \bar z^{\bar j}},
    \qquad
    \frac{d\bar z^{\bar i}}{dt}
    =
   - g^{\bar i j}
    \frac{\partial S_\lambda}{\partial z^j}.
    \label{Eq:DownwardFlowHolo}
\end{equation}
Along such a flow line, one has
\begin{equation}
    \frac{dS_\lambda}{dt}
    =
    \frac{\partial S_\lambda}{\partial z^i}
    \frac{dz^i}{dt}
    =
    -g^{i\bar j}
    \frac{\partial S_\lambda}{\partial z^i}
    \frac{\partial \overline{S_\lambda}}{\partial \bar z^{\bar j}}.
\end{equation}
resulting in the conservation law
\begin{equation}
    \dv{}{t} \, \Im S_\lambda = 0,
    \label{Eq:ConservationLaw}
\end{equation}
and 
\begin{equation}
    \dv{}{t}\, \Re S_\lambda \leq 0.
\end{equation}
In other words, the downward flow decreases $\Re S_\lambda$ while keeping $\Im S_\lambda$ fixed. Moving away from the saddle along a Lefschetz thimble corresponds to following this flow backwards: $\Re S_\lambda$ then increases, so the phase of $e^{-S_\lambda}$ remains constant while its absolute value
decreases. This will be at the heart of Stokes phenomena. Similarly, the upward flow is obtained by flipping the sign in Eq. \eqref{Eq:DownwardFlowHolo} and one gets
\begin{equation}
    \frac{dz^i}{dt}
    =
    g^{i\bar j}
    \frac{\partial \overline{S_\lambda}}{\partial \bar z^{\bar j}},
    \qquad
    \frac{d\bar z^{\bar i}}{dt}
    =
    g^{\bar i j}
    \frac{\partial S_\lambda}{\partial z^j}.
    \label{Eq:UpwardFlowHolo}
\end{equation}

\subsection{Lefschetz Thimbles}
Let $p_\sigma$ be a non-degenerate critical point of $S_\lambda$. The Lefschetz thimble $\J_\sigma$ associated with $p_\sigma$ is defined as the set of points obtained by downward flow starting from $p_\sigma$. More precisely, this is the set of endpoints at some fixed flow time, that we take to be $t=0$, of solutions to Eq. \eqref{Eq:DownwardFlowHolo} satisfying
\begin{equation}
    \lim_{t\to\infty} z(t) = p_\sigma.
\end{equation}
This boils down to say that $\J_\sigma$ is the stable manifold of $p_\sigma$ for the downward flow of $\Re S_\lambda$. Since the Morse index is $n$, the thimble $\J_\sigma$ is a real $n$-dimensional submanifold of the complex $n$-fold $\M$. Similarly, we define the dual thimble $\K_\sigma$ using upward flow Eq. \eqref{Eq:UpwardFlowHolo}, consisting of the points lying on upward-flow trajectories emanating from $p_\sigma$ at $t\to \infty$. Equivalently, $\K_\sigma$ is the unstable manifold of $p_\sigma$ with respect to the downward flow. The two families of cycles have opposite asymptotic behaviour. Along $\J_\sigma$, the function $h$ increases away from the critical point and the cycle ends at infinity in regions where $h \to \infty$. This makes
\begin{equation}
    \int_{\J_\sigma} dz\, e^{-S_\lambda}
\end{equation}
convergent. On the other hand, along $\K_\sigma$, $h$ decreases away from the critical point, so that $\K_\sigma$ ends at infinity in regions where $h\to -\infty$, making them bad integration cycles for $e^{-S_\lambda}$. Instead, we will see that they allow us to compute intersection numbers $n_\sigma$. Before going there, note that the conservation of $\Im S_\lambda$ along the flow implies that each thimble $\J_\sigma$ lies on the level set 
\begin{equation}
    \Im S_\lambda = \Im S_\lambda(p_\sigma) \qquad \Leftrightarrow \qquad e^{-S_\lambda} = e^{-i\, \Im S_\lambda(p_\sigma)}e^{-h},
\end{equation}
such that the phase is constant on the thimble. This is the steepest-descent property of Lefschetz thimbles.\\

\paragraph{Relative Homology and Convergence at Infinity.}
Since the relevant cycles for the integral of $e^{-S_\lambda}$ have ends at infinity in regions where $h=\Re S_\lambda$ is large and positive, ordinary homology is not the appropriate language for the
integration cycles \cite{witten2010analyticcontinuationchernsimonstheory}. These ends are treated as allowed boundaries, so the natural objects
are relative homology classes.

Choose and fix $T>0$ sufficiently large so that $\pm T$ are regular values of $h$ and no relevant critical value lies beyond the
corresponding level sets. We define the region of good asymptotic behaviour by
\begin{equation}
    \M^{T}
    =
    \left\{
        z\in\M\,\middle|\,h(z)>T
    \right\}.
\end{equation}
An admissible integration cycle is allowed to have its boundary at infinity inside $\M^{T}$. Hence, the Lefschetz thimbles define relative
homology classes
\begin{equation}
    [\J_\sigma]\in H_n(\M,\M^{T};\mathbb Z).
\end{equation}
The dual cycles instead extend towards regions where $h$ is large and negative. Defining
\begin{equation}
    \M_{-T}
    =
    \left\{
        z\in\M\,\middle|\,h(z)<-T
    \right\},
\end{equation}
the dual thimbles determine relative homology classes
\begin{equation}
    [\K_\sigma]\in H_n(\M,\M_{-T};\mathbb Z).
\end{equation}
Let us emphasise a dimension-counting point. For a general real Morse function $h:\M \to \mathbb{R}$ on a real $n$-dimensional manifold, we saw that a critical point of Morse index $k$ has a downward-flow cycle of dimension $n-k$ and an upward-flow dual cycle of complementary dimension $k$. Thus, in the non-compact setting, these cycles naturally define classes in
\begin{equation}
    H_{n-k}(\M,\M^{T};\mathbb{Z})
    \qquad \text{and} \qquad
    H_{k}(\M,\M_{-T};\mathbb{Z}),
\end{equation}
respectively. In the holomorphic case, however, $\M$ is a complex $n$-fold, hence has real dimension $2n$, and $h=\Re S_\lambda$ has Morse index $n$ at every non-degenerate critical point of $S_\lambda$. Consequently both the Lefschetz thimbles and their dual upward thimbles are middle-dimensional real $n$-cycles. \\

At this stage, we need a topological way of determining which thimbles contribute to a given integration cycle. Indeed, the Lefschetz thimbles $\J_\sigma$ form a basis of the convergent relative homology group, so an admissible contour $\C$ can be expanded as
\begin{equation}
    \C = \sum_\sigma n_\sigma \J_\sigma .
    \label{Eq:IntegrationContour}
\end{equation}
The goal is then to compute the integer coefficients $n_\sigma$, and which is precisely why we introduced the dual
thimbles $\K_\sigma$. The intersection pairing 
\begin{equation}
    \langle\, \cdot\, , \,\cdot\,  \rangle \, : \, H_n(\M,\M^{T};\mathbb Z)\times H_n(\M,\M_{-T};\mathbb Z) \to \mathbb Z
\end{equation}
tells us how the original integration cycle decomposes into convergent steepest-descent cycles. If there is no flow between two distinct critical points, one can choose the thimbles so that they satisfy
\begin{equation}
    \langle \J_\sigma , \K_\tau \rangle = \delta_{\sigma \tau}.
    \label{Eq:ThimblesIntersection}
\end{equation}
With suitable orientation conventions, the local intersection of $\J_\sigma$ and $\K_\sigma$ at $p_\sigma$ contributes as $+1$. Consequently, if an admissible contour $\C$ defines a class in $H_n(\M,\M^{T};\mathbb Z)$, then it admits the expansion Eq. \eqref{Eq:IntegrationContour} where the coefficients are obtained by intersecting with the dual thimbles
\begin{equation}
    n_{\sigma} = \langle \, \C, \K_\sigma \,\rangle .
    \label{Eq:IntersectionFormula}
\end{equation}
Finally, we can write the original oscillatory integral Eq. \eqref{Eq:OscillatoryIntegral} as
\begin{equation}
    Z = \sum_\sigma n_\sigma \int_{\J_\sigma} dz\, e^{-S_\lambda(z)}.
\end{equation}
This is called the \textit{Picard--Lefschetz decomposition}. In many applications\footnote{See \textit{e.g.} \cite{witten2010analyticcontinuationchernsimonstheory} for a detailed introductory example (the Airy Function).}, the original contour is not itself strictly contained in the interior of a good asymptotic region. It might lie along the boundary between two sectors of convergence. In such cases, one can deform slightly the original contour into an admissible one, and then use the intersection formula Eq. \eqref{Eq:IntersectionFormula} to determine which thimbles contribute to the original integration contour.

\subsection{Stokes Walls and Learning How to Jump}
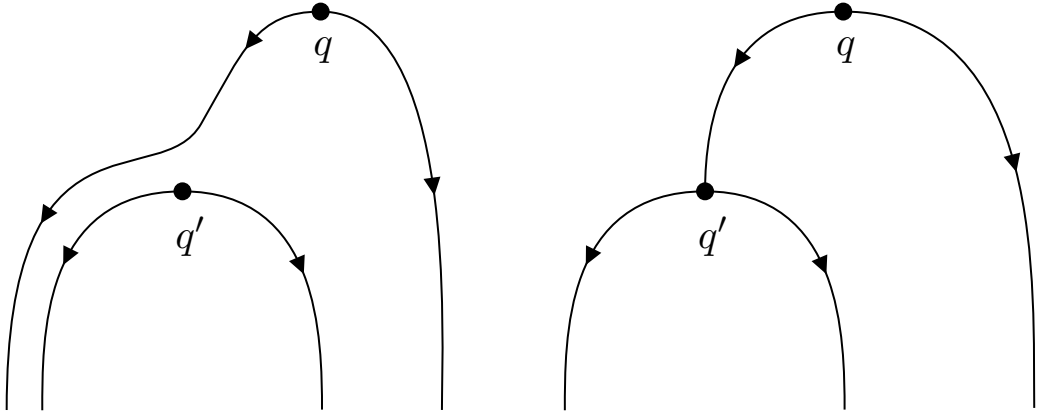
\begin{figure}
    \centering

    \noindent
\begin{minipage}{0.49\linewidth}\centering
\tikzset{every picture/.style={line width=0.75pt}} %set default line width to 0.75pt        
\begin{tikzpicture}[x=0.75pt,y=0.75pt,yscale=-1,xscale=1]

%Shape: Circle [id:dp7153752844173473] 
\draw  [fill={rgb, 255:red, 0; green, 0; blue, 0 }  ,fill opacity=1 ] (166.33,120) .. controls (166.33,117.79) and (168.12,116) .. (170.33,116) .. controls (172.54,116) and (174.33,117.79) .. (174.33,120) .. controls (174.33,122.21) and (172.54,124) .. (170.33,124) .. controls (168.12,124) and (166.33,122.21) .. (166.33,120) -- cycle ;
%Curve Lines [id:da9372013562391058] 
\draw    (170.33,120) .. controls (239.33,120.17) and (240.33,194.17) .. (240.33,229.5) ;
\draw [shift={(231.21,161.56)}, rotate = 246.98] [fill={rgb, 255:red, 0; green, 0; blue, 0 }  ][line width=0.08]  [draw opacity=0] (8.93,-4.29) -- (0,0) -- (8.93,4.29) -- cycle    ;
%Curve Lines [id:da04340654051399684] 
\draw    (170.33,120) .. controls (99.33,120.17) and (100,194.5) .. (100,229.83) ;
\draw [shift={(110.5,156.96)}, rotate = 296.22] [fill={rgb, 255:red, 0; green, 0; blue, 0 }  ][line width=0.08]  [draw opacity=0] (8.93,-4.29) -- (0,0) -- (8.93,4.29) -- cycle    ;
%Shape: Circle [id:dp5414068392004774] 
\draw  [fill={rgb, 255:red, 0; green, 0; blue, 0 }  ,fill opacity=1 ] (235.67,29.83) .. controls (235.67,27.62) and (237.46,25.83) .. (239.67,25.83) .. controls (241.88,25.83) and (243.67,27.62) .. (243.67,29.83) .. controls (243.67,32.04) and (241.88,33.83) .. (239.67,33.83) .. controls (237.46,33.83) and (235.67,32.04) .. (235.67,29.83) -- cycle ;
%Curve Lines [id:da20061756024875432] 
\draw    (239.67,29.83) .. controls (308.67,30) and (300.5,194.33) .. (300.5,229.67) ;
\draw [shift={(296.71,122.04)}, rotate = 262.22] [fill={rgb, 255:red, 0; green, 0; blue, 0 }  ][line width=0.08]  [draw opacity=0] (8.93,-4.29) -- (0,0) -- (8.93,4.29) -- cycle    ;
%Curve Lines [id:da21940881719768845] 
\draw    (239.67,29.83) .. controls (207,29.83) and (200.5,50) .. (179.67,86.17) .. controls (158.83,122.33) and (82.14,72.29) .. (82.14,229.71) ;
\draw [shift={(201.81,48.72)}, rotate = 308.94] [fill={rgb, 255:red, 0; green, 0; blue, 0 }  ][line width=0.08]  [draw opacity=0] (8.93,-4.29) -- (0,0) -- (8.93,4.29) -- cycle    ;
\draw [shift={(99.27,136.12)}, rotate = 302.17] [fill={rgb, 255:red, 0; green, 0; blue, 0 }  ][line width=0.08]  [draw opacity=0] (8.93,-4.29) -- (0,0) -- (8.93,4.29) -- cycle    ;

% Text Node
\draw (164.8,130.4) node [anchor=north west][inner sep=0.75pt]  [xscale=1.4,yscale=1.4]  {$q'$};
% Text Node
\draw (234,40.8) node [anchor=north west][inner sep=0.75pt]  [xscale=1.4,yscale=1.4]  {$q$};

\end{tikzpicture}
\end{minipage}\hfill
\begin{minipage}{0.49\linewidth}\centering
\tikzset{every picture/.style={line width=0.75pt}} %set default line width to 0.75pt        
\begin{tikzpicture}[x=0.75pt,y=0.75pt,yscale=-1,xscale=1]

%Shape: Circle [id:dp3595864929215711] 
\draw  [fill={rgb, 255:red, 0; green, 0; blue, 0 }  ,fill opacity=1 ] (240.86,122.6) .. controls (240.86,120.39) and (242.65,118.6) .. (244.86,118.6) .. controls (247.07,118.6) and (248.86,120.39) .. (248.86,122.6) .. controls (248.86,124.81) and (247.07,126.6) .. (244.86,126.6) .. controls (242.65,126.6) and (240.86,124.81) .. (240.86,122.6) -- cycle ;
%Curve Lines [id:da5930007124934447] 
\draw    (244.86,122.6) .. controls (313.86,122.77) and (314.86,196.77) .. (314.86,232.1) ;
\draw [shift={(305.73,164.16)}, rotate = 246.98] [fill={rgb, 255:red, 0; green, 0; blue, 0 }  ][line width=0.08]  [draw opacity=0] (8.93,-4.29) -- (0,0) -- (8.93,4.29) -- cycle    ;
%Curve Lines [id:da7349898220966213] 
\draw    (244.86,122.6) .. controls (173.86,122.77) and (174.52,197.1) .. (174.52,232.43) ;
\draw [shift={(185.03,159.56)}, rotate = 296.22] [fill={rgb, 255:red, 0; green, 0; blue, 0 }  ][line width=0.08]  [draw opacity=0] (8.93,-4.29) -- (0,0) -- (8.93,4.29) -- cycle    ;
%Shape: Circle [id:dp9774366703925483] 
\draw  [fill={rgb, 255:red, 0; green, 0; blue, 0 }  ,fill opacity=1 ] (310.19,32.43) .. controls (310.19,30.22) and (311.98,28.43) .. (314.19,28.43) .. controls (316.4,28.43) and (318.19,30.22) .. (318.19,32.43) .. controls (318.19,34.64) and (316.4,36.43) .. (314.19,36.43) .. controls (311.98,36.43) and (310.19,34.64) .. (310.19,32.43) -- cycle ;
%Curve Lines [id:da25468420837133177] 
\draw    (314.19,32.43) .. controls (417,32.75) and (409.33,173.58) .. (410,231.25) ;
\draw [shift={(401.04,112.89)}, rotate = 255.45] [fill={rgb, 255:red, 0; green, 0; blue, 0 }  ][line width=0.08]  [draw opacity=0] (8.93,-4.29) -- (0,0) -- (8.93,4.29) -- cycle    ;
%Curve Lines [id:da03158168370426817] 
\draw    (314.19,32.43) .. controls (254.67,32.5) and (244.86,87.27) .. (244.86,122.6) ;
\draw [shift={(259.39,60.56)}, rotate = 304.52] [fill={rgb, 255:red, 0; green, 0; blue, 0 }  ][line width=0.08]  [draw opacity=0] (8.93,-4.29) -- (0,0) -- (8.93,4.29) -- cycle    ;

% Text Node
\draw (239.32,133) node [anchor=north west][inner sep=0.75pt]  [xscale=1.4,yscale=1.4]  {$q'$};
% Text Node
\draw (308.52,43.4) node [anchor=north west][inner sep=0.75pt]  [xscale=1.4,yscale=1.4]  {$q$};

\end{tikzpicture}
\end{minipage}

    \caption{Schematic representation of a Stokes jump \cite{witten2010analyticcontinuationchernsimonstheory}. When two critical points $q$ and $q'$ satisfy $\Im S(q)=\Im S(q')$, a downward-flow trajectory
    may connect them. If $\Re S(q)>\Re S(q')$, crossing the corresponding Stokes wall changes the thimble associated with $q$ by $\mathcal{J}_q \mapsto \mathcal{J}_q \pm \mathcal{J}_{q'}$.}
    \label{Fig:StokesJump}
\end{figure}

Eq. \eqref{Eq:ThimblesIntersection} is valid whenever there is no flow between two distinct critical points. When such a flow appears, it means that we are hitting a \textit{Stokes wall}. Let $S_\lambda$ vary with respect to the parameter $\lambda$, such that $S_\lambda=S(z,\lambda)$. The critical points $p_\sigma=p_\sigma(\lambda)$, and hence the corresponding critical values 
\begin{equation}
    S_\sigma(\lambda) = S(p_\sigma(\lambda),\lambda)
\end{equation}
also vary with $\lambda$. The conservation law Eq. \eqref{Eq:ConservationLaw} implies that a gradient-flow trajectory can connect two critical points $p_\sigma$ and $p_\tau$ only if the imaginary part of their critical values agree, that is
\begin{equation}
    \Im S_\sigma(\lambda) = \Im S_\tau (\lambda) \qquad \Leftrightarrow \qquad \Im \left(S_\sigma (\lambda) - S_\tau (\lambda)\right) =0.
\end{equation}
These loci are candidate \textit{Stokes walls}. In particular, for a single complex parameter, it is usual to call them \textit{Stokes rays}. An actual Picard--Lefschetz jump occurs when a gradient-flow trajectory connecting the corresponding critical points exists. Let $q$ and $q'$ denote two such critical points, with Morse values $h(q)$ and $h(q')$. Because $h$ decreases along downward-flow trajectories, a flow line may run from the critical point with larger $h$ to the one with smaller $h$. On the Stokes wall, such connecting trajectories can appear, and the thimble basis becomes ambiguous. A downward trajectory emanating from $q$ may pass through $q'$ and the thimble basis changes by a Picard--Lefschetz transformation
\begin{equation}
    \J_q \mapsto \J_q \pm \J_{q'}, \qquad \J_{q'}\mapsto \J_{q'},
    \label{Eq:StokesJump}
\end{equation}
where the sign depends on the orientation conventions and on the direction in which the Stokes wall is crossed, see Fig. \ref{Fig:StokesJump}. It is important to distinguish the jump of the thimble basis from the behaviour of the integral itself. The original integration cycle $\C$ is a fixed relative homology class, so the value of the integral does not jump. What changes is the expansion of $\C$ in the thimble basis. If the basis changes as in \eqref{Eq:StokesJump}, then the coefficients $n_\sigma$ change in the inverse way, so that the total cycle $\C$, and hence the integral $Z$, remain unchanged.\\

\paragraph{Gamma-function Example.}
The Gamma function provides a simple one-dimensional illustration of this mechanism \cite{Harlow_2011}. Starting from Euler's integral representation and setting $t=\kappa e^z$, one obtains
\begin{equation}
\Gamma(\kappa)
=
\kappa^\kappa
\int_{\C_\Gamma} dz\,
e^{-S_\Gamma(z)},
\qquad
S_\Gamma(z)=\kappa(e^z-z).
\end{equation}
For positive real $\kappa$, the contour $\C_\Gamma=\mathbb{R}$ coincides with the principal thimble $\J_0$. The critical points are
\begin{equation}
z_n=2\pi i n,
\qquad n\in\mathbb Z,
\end{equation}
and the corresponding critical values are $S_\Gamma(z_n)=\kappa(1-2\pi i n)$. The Stokes wall lies at $\Re\kappa=0$. Continuing $\kappa$ through the upper half-plane into the region $\Re\kappa<0$, $\Im\kappa>0$, the integration cycle decomposes as
\begin{equation}
[\C_\Gamma]
=
\sum_{n=0}^{\infty}[\J_n].
\end{equation}
Since adjacent thimbles are related by $z\mapsto z+2\pi i$, their contributions differ by a factor $e^{2\pi i\kappa}$, and the infinite sum generates the geometric factor
\begin{equation}
\sum_{n=0}^{\infty}e^{2\pi i n\kappa}
=
\frac{1}{1-e^{2\pi i\kappa}}.
\end{equation}
The reciprocal Gamma function instead admits the Hankel representation
\begin{equation}
\frac{1}{\Gamma(\kappa)}
=
\frac{1}{2\pi i}
\int_{\H_t}dt\,t^{-\kappa}e^t
=
-\frac{\kappa^{1-\kappa}}{2\pi i}
\int_{\C_{1/\Gamma}}dz\,
e^{\kappa(e^{-z}+z)}e^{-z},
\end{equation}
where $\H_t$ encircles the negative real axis in the $t$-plane and $\C_{1/\Gamma}$ is its image under $t=\kappa e^{-z}$, with the induced orientation. For the same continuation into the upper-left quadrant, its thimble decomposition is
\begin{equation}
[\C_{1/\Gamma}]
=
[\widetilde{\J}_0]-[\widetilde{\J}_1].
\end{equation}
The relative minus sign produces the factor $1-e^{2\pi i\kappa}$. This two-thimble structure is the finite-dimensional prototype of the analogous factor appearing in the timelike Liouville partition function.

\section{Saddle Structure in Timelike Liouville}
\label{Sec:SaddleContribtL}
We now combine the review of timelike Liouville theory in Sec. \ref{Sec:TimelikeLiouville} with the Picard-Lefschetz framework of Sec. \ref{Sec:MorseTheory} to analyse the saddle contribution to the sphere partition function. We start from the action in Eq. \eqref{Eq:tLAction}, whose classical equation of motion is
\begin{equation}
    2 \widetilde \nabla^2 \phi - \frac{2q}{v} + 8\pi \beta \Lambda  e^{2\beta \phi}=0.
    \label{Eq:EOMClassical}
\end{equation}
We restrict ourselves to a two-sphere of area $4\pi v$, with fiducial metric
\begin{equation}
    d\tilde s^2 = \frac{4v}{(1+z\barz)^2}dz d\barz,
\end{equation}
and without loss of generality we stick to $\beta>0$ as $q(-\beta)=-q(\beta)$ and the simultaneous transformation $(\beta,\phi)\mapsto (-\beta,-\phi)$ leaves the action invariant. This equivalence also maps the integration cycle. The equation of motion admits a set of constant solutions given by
\begin{equation}
    \varphi_n =\frac{1}{2\beta} \log \left(\frac{q}{4 \pi  \beta  \Lambda  v}\right) + i \frac{n\pi}{\beta} , \qquad \forall \, n\in \mathbb Z.
    \label{Eq:Saddles}
\end{equation}
We are interested in understanding  saddle-point approximation to the genus-zero timelike Liouville path integral Eq. \eqref{Eq:tLPI}. To isolate the integration-cycle and saddle-content questions, we study the following zero-mode model
\begin{equation}
    \Z_{\rm saddle} (\Lambda,\beta) = \int d\phi \, e^{-\S_{\rm saddle}[\phi]}
\end{equation}
where the contour of integration is intentionally left implicit at this stage, and
\begin{equation}
    S(\phi,\beta) = \S_{\rm saddle}[\phi] = 4 \pi  \Lambda  v e^{2 \beta  \phi }-2 q \phi,
\end{equation}
with $q=\beta^{-1}-\beta$. The non-zero modes provide the perturbative fluctuation factor common to the saddle sectors considered below. Letting
\begin{equation}
    t = 4\pi v \Lambda e^{2\beta\phi},
\end{equation}
one can recover the ``Gamma representation"
\begin{align}
    \Z_{\mathcal C} &= \int_{\mathcal C} d\phi \, e^{-4\pi v \Lambda e^{2\beta\phi}+2q\phi}\nonumber\\
    &= \frac{(4\pi v \Lambda)^{-\lambda}}{2\beta} \int_{\mathcal C_t} dt\, t^{\lambda-1} e^{-t},
\end{align}
where the contour of integration $\mathcal C$ will be discussed in what follows and $\lambda = q/\beta$.\\

\subsection{Morse Theory Dictionary}
Our Morse function is defined as
\begin{equation}
    h(\phi,\beta) = \Re S(\phi,\beta).
\end{equation}
The critical points are $\phi=\varphi_n$, for $n\in\mathbb Z$. At these points, the values of $S$ and $h$ are
\begin{align}
    S_n &= \lambda \left(1-\log \left(\frac{\lambda}{4 \pi  \Lambda  v}\right)-2\pi i n\right),\\
    h_n &= \lambda\left(1-\log\left|\frac{\lambda}{4\pi v  \Lambda}\right|\right).
\end{align}
They are non-degenerate because
\begin{equation}
    \eval{\pdvsq{S}{\phi}}{\phi=\varphi_n} = 4 \left(1-\beta ^2\right) \neq 0
\end{equation}
for $\beta\neq1$. At the particular value $\beta=1$ finite critical points escape to infinity and the finite-saddle Picard--Lefschetz analysis breaks down. Related field-theoretic aspects of this point have been studied in \cite{Ribault:2015sxa,Gutperle_2003}.

\paragraph{Local Analysis.}
Before solving the full gradient-flow equations, it is useful to determine the local tangent directions of the thimbles near each critical point. These directions are fixed by the quadratic expansion of the holomorphic function $S(\phi,\beta)$ around the saddle and provide the initial data for the global Picard-Lefschetz flows. We seek the steepest-ascent directions, along which $h$ increases most rapidly away from the saddle. The steepest descent directions are obtained by reversing the sign. Near a saddle, we can expand $S$ around $\varphi_n$ as
\begin{equation}  S(\phi,\beta)=S(\varphi_n,\beta)+\frac12(\phi-\varphi_n)^2 \eval{\pdvsq{S(\phi,\beta)}{\phi}}{\phi = \varphi_n}+\mathcal{O}\left((\phi-\varphi_n)^3\right),
\end{equation}
and the descent/ascent directions are determined by the phase of the second derivative. The Morse function is thus
\begin{align}
   h(\phi,\beta) &= h_n + \frac12 r_n^2 |S''_n|\cos(\alpha_n+2\theta_n),
\end{align}
with 
\begin{equation}
    \phi-\varphi_n = r_n e^{i\theta_n}
\end{equation}
and
\begin{equation}
   \eval{\pdvsq{S(\phi,\beta)}{\phi}}{\phi = \varphi_n} \equiv S''_n= |S''_n|e^{i\alpha_n}.
\end{equation}
The condition for steepest ascent is
\begin{equation}
    \cos(\alpha_n+2\theta_n) = 1 \quad \Leftrightarrow \quad \alpha_n+2\theta_n = - 2k \pi, \quad k\in\mathbb{Z}.
\end{equation}
The condition for the steepest descent is 
\begin{equation}
    \cos(\alpha_n+2\theta_n) = -1 \quad \Leftrightarrow \quad \alpha_n+2\theta_n = \pi + 2k \pi, \quad k\in\mathbb{Z}.
\end{equation}
The rays are therefore given by
\begin{equation}
    {\rm arg}(\phi-\varphi_n)
    =
    \theta_n
    =
    -\frac12\arg S''_n+k\pi,
    \qquad k\in\mathbb Z.
\end{equation}
for the steepest-ascent directions. Similarly, the steepest-descent condition is
\begin{equation}
    {\rm arg}(\phi-\varphi_n)
    =
    \theta_n
    =
    \frac12\left(\pi-\arg S''_n\right)+k\pi,
    \qquad k\in\mathbb Z,
\end{equation}\\

\paragraph{Downward Flow Equations.} 
We now study the global representatives of the thimbles. We choose the
flat Kähler metric
\begin{equation}
ds^2=d\phi\,d\bar\phi
\end{equation}
as an auxiliary metric on the complexified zero-mode space
$\M=\mathbb C_\phi$. The auxiliary metric selects convenient representatives of the stable and unstable manifolds, while their relative homology classes remain unchanged under smooth metric deformations, provided that no saddle connection appears and no asymptotic boundary condition changes. At a Stokes ray, flow lines connecting
distinct critical points can appear, and the thimble basis can jump. The flow equations are
\begin{equation}
    \dv \phi t =  - \pdv{\bar S}{\bar \phi}, \qquad \dv{\bar\phi} {t} = - \pdv{ S}{ \phi}.
\end{equation}
The conservation law is
\begin{equation}
    \dv{h}{t} \leq 0,\qquad \dv{}{t}\Im S = 0.
    \label{Eq:ConservationLawtL}
\end{equation}\\

\subsection{Embroidering the Lefschetz Thimbles} 
The Lefschetz thimble $\J_n$ associated with $\varphi_n$ is the set of points satisfying
\begin{equation}
    \lim_{t\to\infty} \phi(t) = \varphi_n.
\end{equation}
By construction, $\J_n$ is a steepest-ascent cycle of $h$, equivalently a steepest-descent contour for the integrand $e^{-S}$. Hence,
\begin{equation}
    \int_{\J_n} d\phi\, e^{-S(\phi,\beta)} = e^{-S_n}\int_{\J_n} d\phi\, e^{-S(\phi,\beta)+S_n}
\end{equation}
converges. Let 
\begin{equation}
    \lambda = \frac q \beta, \qquad \xi = 2\beta (\phi - \varphi_n)= x+i y,
\end{equation}
for $x,y\in \mathbb R$. Using
\begin{equation}
    \eval{\pdv{S(\phi)}{\phi}}{\phi=\varphi_n} = 0 ,
\end{equation}
one finds
\begin{equation}
    \lambda = 4\pi v \Lambda e^{2\beta\varphi_n},
\end{equation}
such that the exponent near the saddle can be written as
\begin{equation}
    S(\phi,\beta)-S_n = \lambda \left( e^\xi -\xi -1\right).
\end{equation}
Therefore, this gives the following decomposition into real and imaginary parts
\begin{align}
    \mathfrak h(x,y) &= h(\phi,\beta) - h_n = \lambda(e^x \cos y -x-1),\label{Eq:RealPart}\\
    \mathfrak S(x,y) &= \Im S(\phi,\beta) - \Im S_n = \lambda(e^x \sin y-y).
\end{align}
Because of the conservation law Eq. \eqref{Eq:ConservationLawtL}, the thimble attached to a given critical point lies on the level set
\begin{equation}
    \mathfrak S(x,y) = 0 \quad \Leftrightarrow \quad e^x \sin y - y=0.
\end{equation}
Requiring this level set to pass through the critical point at $\xi=0$, one obtains two branches. One consists of simple straight lines
\begin{align}
    \H_n &= \left\{(x,y) \, \big| \, y=0 , \, x\in \mathbb R \right\}\\
    &=\left\{\phi\, \big| \, \phi = \varphi_n + \frac{x}{2\beta} , \, \forall x\in \mathbb R \right\},
\end{align}
while the other one is given by
\begin{align}
    \widetilde\H_n &= \left\{(x,y) \, \big| \,  x = \log \left( \frac{y}{\sin y}\right), \, -\pi <y <\pi \right\}\\
    &=\left\{\phi\, \big| \, \phi = \varphi_n + \frac{1}{2\beta} \left[\log\left(\frac{y}{\sin y}\right)+iy\right] , \, \forall y\in (-\pi,\pi) \right\},
\end{align}
where the condition of passing through the saddle follows from $\lim_{y\to 0} \xi = 0$. The two branches have different asymptotic behaviour. Along $\H_n$, one can rewrite Eq. \eqref{Eq:RealPart} as
\begin{equation}
    \mathfrak h(x)=\eval{\mathfrak h (x,0)}{\H_n} = \lambda(e^x -x-1)
\end{equation}
such that 
\begin{equation}
    \lim_{x\to \pm \infty}\mathfrak h(x) = \begin{cases}
        +\infty \qquad \text{if }\lambda>0,\\
        -\infty \qquad \text{if }\lambda<0.
    \end{cases}
\end{equation}
Along $\widetilde \H_n$,  Eq. \eqref{Eq:RealPart} becomes
\begin{equation}
    \tilde{\mathfrak h}(y)=\eval{\mathfrak h\left(\log \left( \frac{y}{\sin y}\right),y\right)}{\widetilde\H_n} =\lambda \left( y \cot y -1-\log\left(\frac{y}{\sin y}\right)\right)
\end{equation}
such that
\begin{equation}
    \lim_{y\to \pm \pi} \tilde{\mathfrak h}(y) = \begin{cases}
        -\infty \qquad \text{if }\lambda>0,\\
        +\infty \qquad \text{if }\lambda<0.
    \end{cases}
\end{equation}
Hence, the sign of $\lambda$ determines which branch is the Lefschetz thimble and which is its dual. For $\beta\in (0,1)$, one has $\lambda>0$, while for $\beta>1$, one has $\lambda<0$, and the roles of the two branches are exchanged.

\subsection{Stokes Decomposition of the Zero-Mode Contour}
\label{Sec:StokesDiscussionBetaOne}
\subsubsection{Integration Contour}
In a region with no Stokes degeneracy, an admissible cycle decomposes as
\begin{equation}
    \C = \sum_{\sigma} n_\sigma \J_\sigma, \qquad  n_{\sigma} = \langle \, \C, \K_\sigma \,\rangle,
\end{equation}
with $\J_\sigma$ the Lefschetz thimbles and $\K_\sigma$ their duals. We must therefore compute the intersection coefficients $n_\sigma$ for the contour of interest. The real-line prescription for the Liouville zero mode is not admissible throughout the parameter space. Indeed, after the change of variables
\begin{equation}
	t=4\pi v\Lambda e^{2\beta\phi},
\end{equation}
the zero-mode integral takes the Gamma-type form
\begin{equation}
	Z_{\rm saddle}\propto
	\int_{\C_t} dt\, t^{\lambda-1}e^{-t},
	\qquad
	\lambda=\frac{q}{\beta}
	=\frac{1}{\beta^2}-1.
    \label{Eq:EulerRep}
\end{equation}

\begin{figure}\hspace{-1cm}
    \centering
    \includegraphics[width=0.8\linewidth]{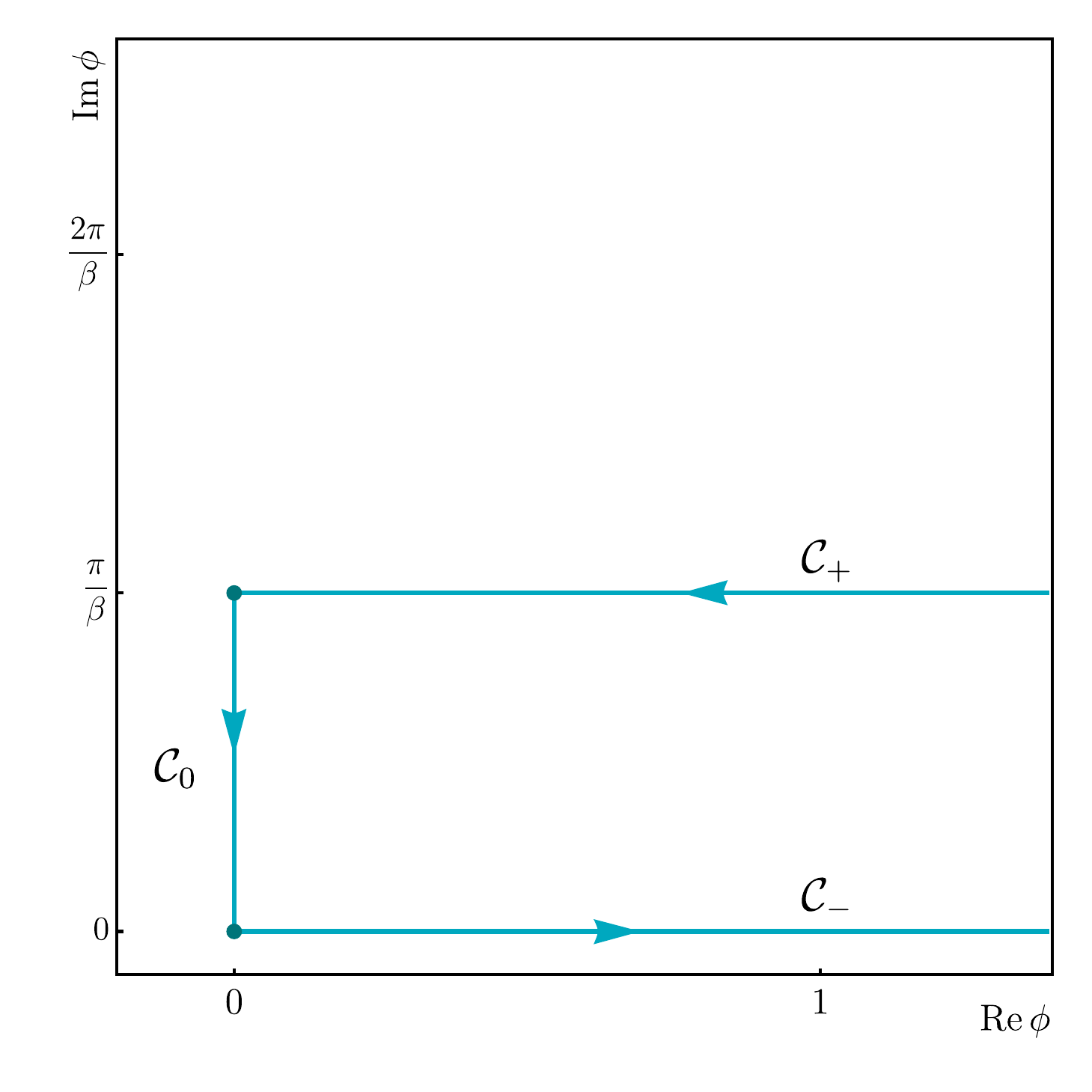}
    \caption{The chosen Hankel contour, for $\beta=0.5$, $v=1$ and $\Lambda=1$.}
    \label{Fig:HankelContour}
\end{figure}

\begin{figure}\hspace{-2cm}
    \centering
    \includegraphics[width=0.9\linewidth]{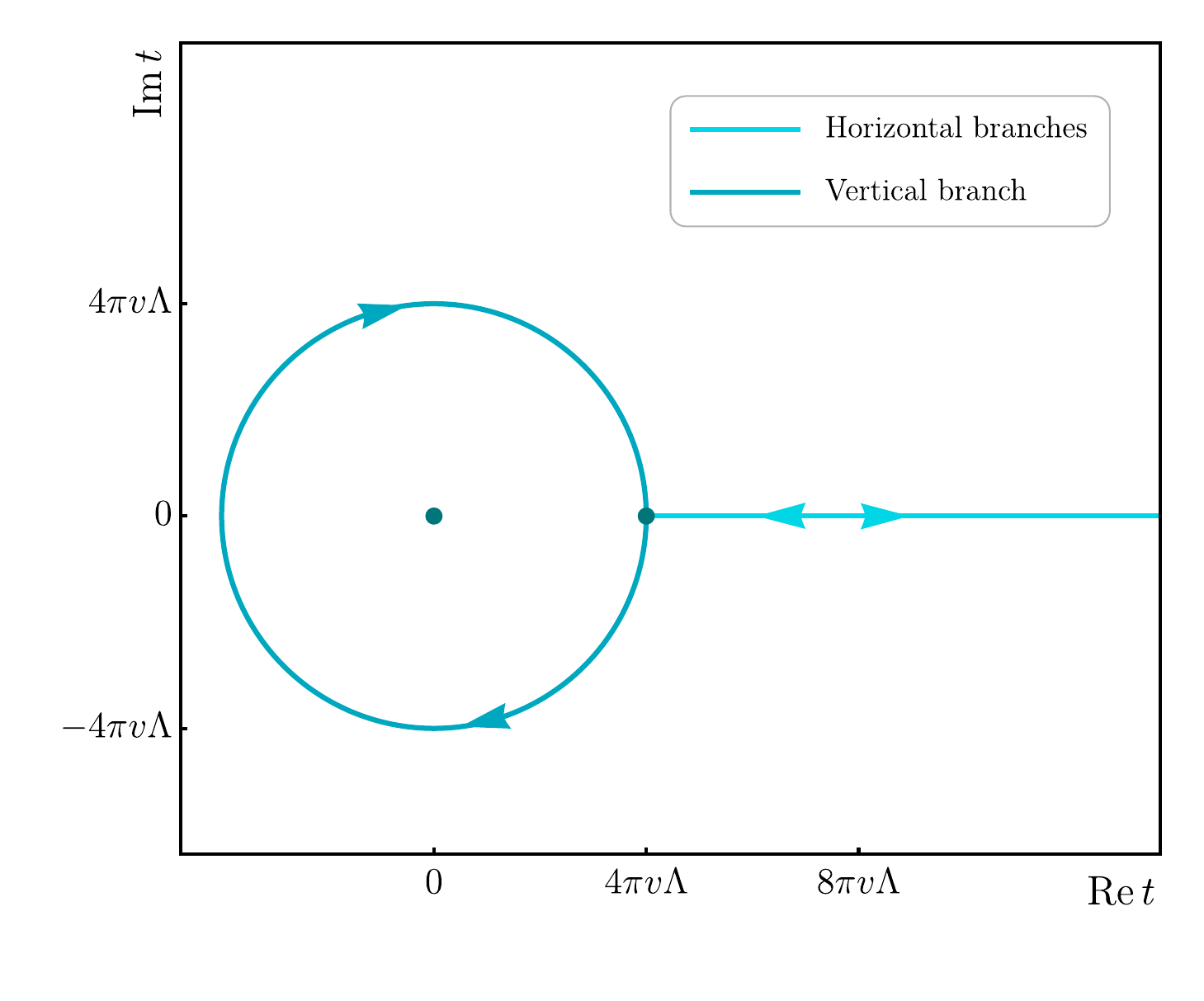}
    \caption{The Hankel contour mapped in the $t$-plane, for $\beta=0.5$, $v=1$ and $\Lambda=1$.}
    \label{Fig:tPlane}
\end{figure}

For $0<\beta<1$, one has $\lambda>0$, and integration along the positive real $t$-axis gives a convergent Euler integral. A Hankel-type cycle is also available, but it represents a different relative homology class and generally gives a different zero-mode
amplitude. For $\beta>1$, one has $\lambda<0$, and the real integral diverges. An additional prescription is then necessary. We choose the Hankel-type contour $C_H$ in both regimes, motivated by the inverse-Gamma representation \cite{Harlow_2011,Cao_2023} and by its relation to
the oscillatory factor in the proposed sphere partition functions.
Related contours for charged correlators
\cite{Usciati:2025cdn,Giribet:2026gao} provide further motivation, but do not establish the zero-point prescription. The analysis below determines the thimble decomposition conditional on this choice. In the complex $\phi$-plane, the contour begins at infinity along the horizontal line $\Im\phi=\pi/\beta$ and runs towards $\phi=i\pi/\beta$. It then follows the imaginary axis down to the origin, before returning to infinity along the positive real axis. More precisely, as depicted in Fig. \ref{Fig:HankelContour},
\begin{equation}
\mathcal C_H
=
\mathcal C_{+}\cup\mathcal C_{0}\cup\mathcal C_{-},
\label{Eq:HankelContour}
\end{equation}
where
\begin{align}
\mathcal C_{+}&=\left\{\phi=x+\frac{i\pi}{\beta}\,|\,
x:+\infty\to 0\right\},\\
\mathcal C_{0}&=\left\{\phi=iy\,|\,
y:\frac{\pi}{\beta}\to 0\right\},\\
\mathcal C_{-}&=\left\{ \phi=x\,|\,
x:0\to+\infty\right\}.
\end{align}
The motivation for this choice is the same as in the inverse-Gamma representation (see Appendix C of \cite{Harlow_2011} for a detailed analysis). In the $t$-plane, this maps to $\C_{H_t}$, as shown in Fig. \ref{Fig:tPlane}.\\ 

We now discuss the two non-degenerate regimes, followed by the critical case $\beta=1$. Case I is the regime relevant to the semi-classical limit, as semi-classicality is reached for $q\to \infty,$ that is $\beta\to 0$. The opposite asymptotic limit lies in Case II, for which $\beta\to\infty$, instead gives $q\to-\infty$ and is related to the former through the $\beta\to1/\beta$ ``duality", accompanied by $q\to-q$. In this sense, the large-$\beta$ regime may be viewed as a dual large-background-charge limit. The saddle organisation found below, however, shows that the $\beta>1$ branch is not obtained from the $0<\beta<1$ branch by a trivial rearrangement of the same real saddles.

\subsubsection[Case I -- \texorpdfstring{$0<\beta<1$}{0<β<1}]
{Case I -- $\boldsymbol{0<\beta<1}$}
For $\lambda>0$, equivalently $0<\beta<1$, the critical points obtained in Eq. \eqref{Eq:Saddles} are
\begin{equation}
    \varphi_n =\frac{1}{2\beta}\log \left(\frac{1-\beta^2}{4 \pi  \beta ^2 \Lambda  v}\right) + i \frac{n\pi}{\beta}, \qquad \forall \, n\in \mathbb Z,
\end{equation}
so that one saddle 
\begin{equation}
    \varphi_0 = \frac{1}{2\beta}\log \left(\frac{1-\beta^2}{4 \pi  \beta ^2 \Lambda  v}\right)
\end{equation}
lies on the real axis.
\begin{figure}[h!]
    \centering
    \includegraphics[width=0.8\linewidth]{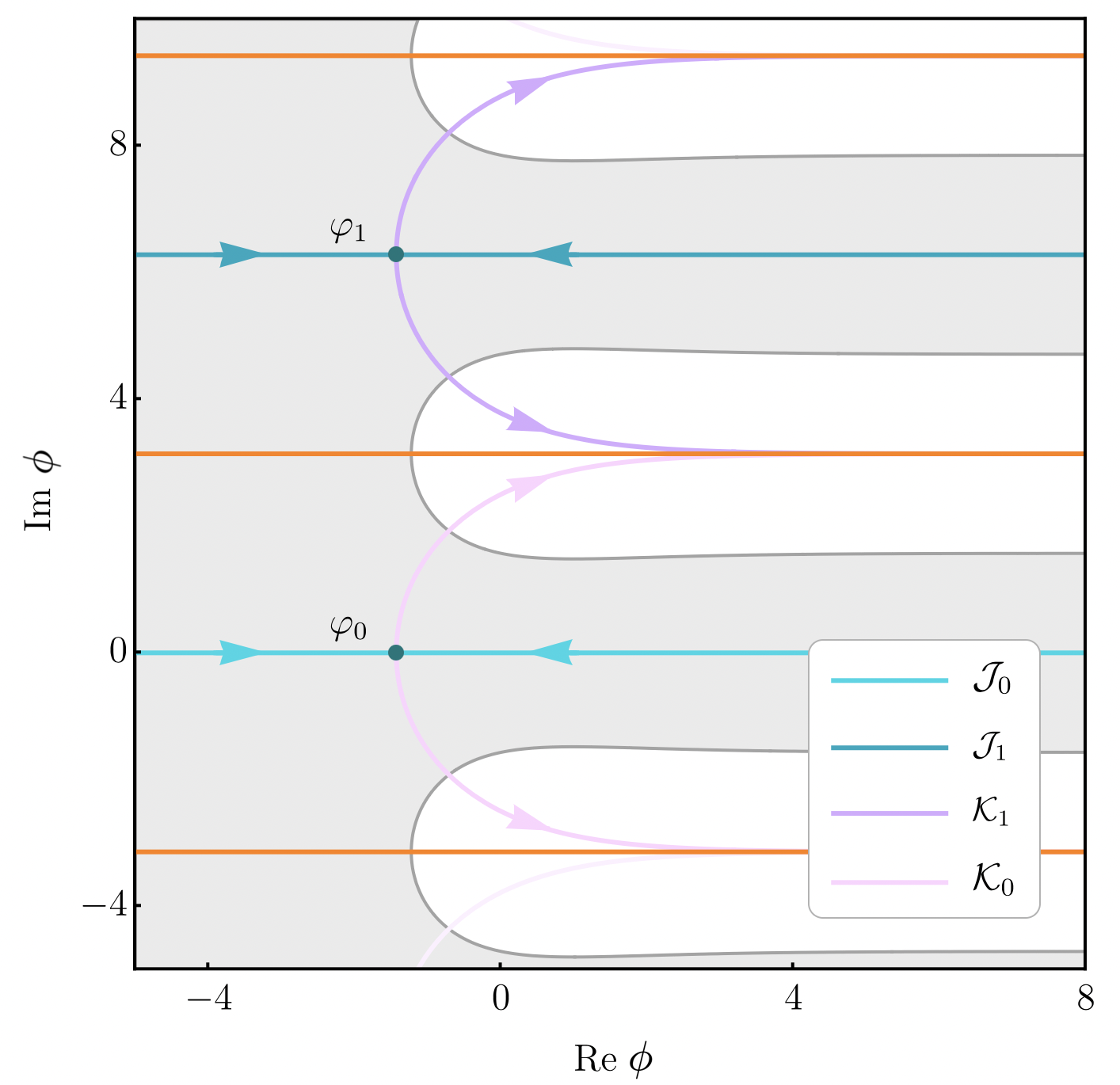}
    \caption{Stokes geometry for $\beta=0.5$, for which we fixed $v=\Lambda=1$. The orange curves are constant-$\Im S$ gradient-flow trajectories. The arrows indicate the direction of downward flow.}
    \label{Fig:BetaLessOne}
\end{figure}
For the Hankel-type contour $\C_H$,  the thimble associated with the real saddle alone is not homologous to the integration contour, as is also apparent from Fig. \ref{Fig:BetaLessOne}. More explicitly, we determine the Lefschetz-thimble decomposition by computing the intersection coefficients introduced in Eq.~\eqref{Eq:IntersectionFormula}. We look at the complex $\phi$-plane as the real two-dimensional smooth manifold $\mathcal M = \mathbb C_\phi \cong \mathbb R^2$ with global coordinates $(\Re \phi, \Im \phi)$, and orient it by
\begin{equation}
    \omega
    =
    d(\Re\phi)\wedge d(\Im\phi),
\end{equation}
such that an ordered basis $(V_1,V_2)$ of $T_p\mathcal M$ is positively
oriented when
\begin{equation}
    \omega_p(V_1,V_2)>0.
\end{equation}
Since each thimble is a connected one-dimensional manifold, its orientation is determined by choosing one of the two tangent directions at a single point and extending this choice continuously along the curve. The Lefschetz thimbles are
\begin{equation}
    \J_n = \left\{\phi\, \big| \, \phi = \varphi_n + \frac{x}{2\beta} , \, \forall x\in \mathbb R \right\}
\end{equation}
and the dual thimbles are
\begin{equation}
    \K_n = \left\{\phi\, \big| \, \phi = \varphi_n + \frac{1}{2\beta} \left[\log\left(\frac{y}{\sin y}\right)+iy\right] , \, \forall y\in (-\pi,\pi) \right\}.
\end{equation}
At the saddle $\varphi_n$, we orient the Lefschetz thimble $\J_n$ so that its positive tangent points in the direction of increasing $\Re\phi$, namely
\begin{equation}
    E_{\J_n}
    =
    \left.
    \frac{\partial}{\partial(\Re\phi)}
    \right|_{\varphi_n}.
\end{equation}
We orient $\K_n$ so that
\begin{equation}
    \omega_{\varphi_n}
    \left(E_{\J_n},E_{\K_n}\right)
    =
    1>0,
\end{equation}
so that the thimbles satisfy the duality convention
\begin{equation}
    \langle\J_m,\K_n\rangle=\delta_{mn}.
\end{equation}
We choose
\begin{equation}
    E_{\K_n}
    =
    \left.
    \frac{\partial}{\partial(\Im\phi)}
    \right|_{\varphi_n}.
\end{equation}
%The arrows displayed in Fig.~\ref{Fig:BetaLessOne} indicate the directions of the gradient flow and not the orientations of the thimbles as one-dimensional relative cycles. In particular, the flow may point away from a saddle along both branches of a dual thimble, whereas the orientation of the thimble is a continuous choice of positive tangent direction along the entire curve.
We now have everything at hand to compute the intersection with the integration cycle
\eqref{Eq:HankelContour}. Since the integrand has no singularity in the finite
$\phi$-plane, we may deform $\C_H$ within the same relative homology class and
choose a representative whose vertical segment lies sufficiently far to the left of
the saddles. For this representative, the vertical segment does not intersect any dual thimble $\K_n$, so that all intersections arise from the two horizontal
components. Along $\K_n$, the parametrised curves $k_n\in \K_n$ are such that
\begin{equation}
    \Im k_n(y)=\frac{n\pi}{\beta}+\frac{y}{2\beta}.
\end{equation}
The lower horizontal component $\C_-$ lies at $\Im\phi=0$, so an intersection
with $\K_n$ requires $y=-2\pi n$. Since $y\in(-\pi,\pi)$, this is only possible
for $n=0$ and $y=0$, corresponding to the saddle $\varphi_0$. Similarly, the
upper horizontal component $\C_+$ lies at $\Im\phi=\pi/\beta$, so an intersection
with $\K_n$ requires $y=2\pi(1-n)$, which is only possible for $n=1$ and $y=0$,
corresponding to $\varphi_1$. Hence,
\begin{equation}
    \langle \C_H,\K_n\rangle=0,
    \qquad \forall n\neq0,1.
\end{equation}\\

We finally need to determine the signs of the two non-vanishing coefficients. With the orientation introduced before, we have that the tangent vectors at the intersection points can be chosen as
\begin{equation}
    \eval{\dot {\mathcal C}_-}{\varphi_0} = \pdv{}{\Re \phi} , \qquad \eval{\dot {\mathcal C}_+}{\varphi_1} = -\pdv{}{\Re \phi}.
\end{equation}
For $n=0$, at $\varphi_0$, one has
\begin{equation}
    \det \begin{pmatrix}
        \,1 & \phantom{-}0 \,\\
        \,0 & \phantom{-}1\,
    \end{pmatrix} = 1 \quad \to \quad \langle\C_H, \K_0\rangle = 1.
\end{equation}
Similarly, for $n=1$, at $\varphi_1$, we have 
\begin{equation}
    \det \begin{pmatrix}
        -1 & \phantom{-}0 \,\\
        \phantom{-}0 & \phantom{-}1\,
    \end{pmatrix} = -1 \quad \to \quad \langle\C_H, \K_1\rangle = -1.
\end{equation}
The relative homology class of the Hankel contour is 
\begin{equation}
    [\mathcal{C}_H] = [\mathcal{J}_0] - [\mathcal{J}_1].
\end{equation}
Under the translation 
\begin{equation}
    \phi \to \phi + i\frac{\pi}{\beta},
\end{equation}
one has
\begin{equation}
    \int_{\J_1} d\phi \, e^{-\S_{\rm saddle}[\phi]} = e^{2\pi i/\beta^2}\int_{\J_0} d\phi \, e^{-\S_{\rm saddle}[\phi]},
\end{equation}
implying that the zero-mode contribution is 
\begin{equation}
    \Z_{\rm saddle}(\Lambda,\beta) = \int_{\C_H} d\phi \, e^{-\S_{\rm saddle}[\phi]} = \left(1-e^{2\pi i/\beta^2}\right)\int_{\J_0} d\phi \, e^{-\S_{\rm saddle}[\phi]}.
\end{equation}
This is precisely the Stokes factor that appears in the timelike DOZZ expression Eq. \eqref{Eq:tLDOZZ}. This statement is contour-dependent. If one restricts to the real contour in the regime where that contour is admissible, the real saddle would be the relevant one. Moreover, it is worth pointing out that, in the introduced $t$ coordinate, one has
\begin{align}
   & \J_0 =\left\{ t\in (0,\infty) \,|\, \, \arg t = 0\right\},\\
    & \J_1 =\left\{ t\in (0,\infty) \,|\, \, \arg t = 2\pi\right\},
\end{align}
such that geometrically, they lie on the same positive real axis, on two different logarithmic sheets. Hence,
\begin{equation}
    \Z_{\J_0} \propto \int_0^\infty dt\, t^{\lambda -1}e^{-t} = \Gamma(\lambda), \qquad \lambda>0.
    \label{Eq:GammaRepBetasmall}
\end{equation}
Therefore, the zero-mode contribution to the two-sphere partition function is 
\begin{equation}
    \Z_{\C_H}(\Lambda,\beta)=\frac{\left(1-e^{2\pi i/\beta^2}\right)\Gamma(1/\beta^2-1)}{2\beta(4\pi v\Lambda)^{1/\beta^2-1}}, \qquad 0<\beta<1.
\end{equation}
This oscillatory factor also appears in the finite-$\beta$ timelike
sphere expression of Giribet and Leoni \cite{Giribet:2022cvw}
\begin{equation}
    \Z_{\rm GL}(\Lambda,\beta)
    =-\frac{(1+\beta^2)
    \left[\pi\Lambda\gamma(-\beta^2)\right]^{-q/\beta}}
    {\pi^3 q\,\gamma(-\beta^2)\gamma(-\beta^{-2})},
    \qquad
    \gamma(x)=\frac{\Gamma(x)}{\Gamma(1-x)}.
\end{equation}
For $0<\beta<1$, choosing
$\arg[\gamma(-\beta^2)]=-\pi$ and applying the reflection formula
rewrites its oscillatory dependence as
$e^{i\pi/\beta^2}\sin(\pi/\beta^2)$, which is proportional to
$1-e^{2\pi i/\beta^2}$. Our analysis identifies the origin of this
factor in the chosen zero-mode contour. \\

\subsubsection[Case II -- \texorpdfstring{$\beta>1$}{β>1}]
{Case II -- $\boldsymbol{\beta>1}$}

\begin{figure}[h]
    \centering
    \includegraphics[width=0.8\linewidth]{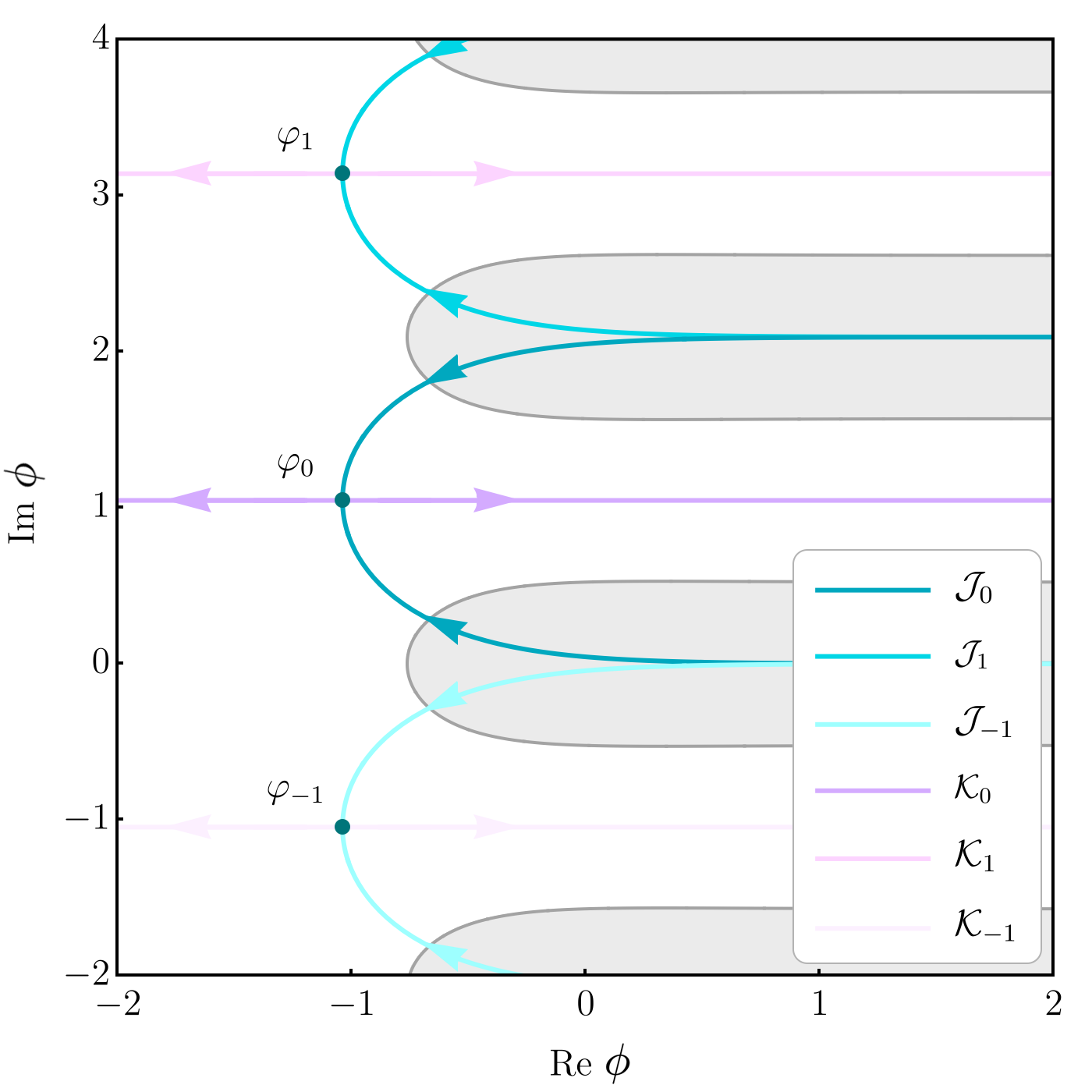}
    \caption{Stokes geometry for $\beta=1.5$, for which we fixed $v=\Lambda=1$. The arrows indicate gradient-flow directions.}
    \label{Fig:BetaBiggerOne}
\end{figure}

We now turn to the regime $\lambda<0$, corresponding to $\beta>1$. The critical points obtained from Eq. \eqref{Eq:Saddles} are
\begin{equation}
\varphi_n
=
\frac{1}{2\beta}
\log\left(
\frac{-\lambda}{4\pi v\Lambda}
\right)
+
i\frac{(2n+1)\pi}{2\beta},
\qquad
\forall \,n\in\mathbb Z.
\end{equation}
Contrary to the previous case, none of the critical points lies on the real axis. In the fundamental strip selected by the Hankel contour
\begin{equation}
0<\Im\phi<\frac{\pi}{\beta},
\end{equation}
there is a unique critical point, namely
\begin{equation}
\varphi_0
=
\frac{1}{2\beta}
\log\left(
\frac{\beta^2-1}{4\pi v\Lambda\beta^2}
\right)
+
i\frac{\pi}{2\beta}.
\end{equation}
The change of sign of $\lambda$ exchanges the roles of the two families of steepest-flow trajectories relative to the regime $0<\beta<1$. The Lefschetz thimbles are now the curved trajectories
\begin{equation}
\J_n
=
\left\{
\phi\,\bigg|\,
\phi
=
\varphi_n
+
\frac{1}{2\beta}
\left[
\log\left(\frac{y}{\sin y}\right)+iy
\right],
\quad
\forall\,y\in(-\pi,\pi)
\right\},
\end{equation}
whereas the dual thimbles are the horizontal lines
\begin{equation}
\K_n
=
\left\{
\phi\,\bigg|\,
\phi
=
\varphi_n+\frac{x}{2\beta},
\quad
\forall\,x\in\mathbb R
\right\}.
\end{equation}
The Lefschetz thimble $\J_0$ associated with the critical point in the fundamental strip interpolates between the two horizontal components of the Hankel contour. Indeed, along $\J_0$ one has
\begin{equation}
\Im\phi
=
\frac{\pi+y}{2\beta},
\qquad
y\in(-\pi,\pi),
\end{equation}
so that its two asymptotic ends approach $\Im\phi = 0$ and $\Im\phi= \pi/\beta$, respectively. This is precisely the pair of asymptotic regions selected by the lower and upper components $\C_-$ and $\C_+$ of the Hankel contour. As is also apparent from Fig.~\ref{Fig:BetaBiggerOne}, the contour $\C_H$ can therefore be continuously deformed onto $\J_0$ within the relevant relative homology class, without crossing any singularity and while preserving the required asymptotic convergence conditions.\\

Equivalently, only the dual thimble $\K_0$ has a non-vanishing intersection number with the Hankel cycle,
\begin{equation}
\langle\C_H,\K_n\rangle
=
\delta_{n0},
\end{equation}
with 
\begin{equation}
    E_{\J_0} = - \pdv{}{\Im \phi}, \quad E_{\K_0} =  \pdv{}{\Re \phi}
\end{equation}
at $\varphi_0$ such that $\mathcal J_0$ is oriented from its upper end toward its lower end, consistently with $\mathcal C_H$, and
\begin{equation}
\langle\C_H,\K_0\rangle=1.
\end{equation}
The Lefschetz-thimble decomposition of the Hankel contour consequently reduces to
\begin{equation}
[\C_H]=[\J_0].
\end{equation}
Accordingly, the zero-mode contribution is given by the single thimble
\begin{equation}
\Z_{\rm saddle}(\Lambda,\beta)
=
\int_{\C_H}d\phi\,e^{-\S_{\rm saddle}[\phi]}
=
\int_{\J_0}d\phi\,e^{-\S_{\rm saddle}[\phi]}.
\end{equation}
In terms of the Gamma function, the thimble $\J_0$ maps to the full Hankel cycle, leading to the inverse-Gamma representation
\begin{align}
\Z_{\rm saddle}(\Lambda,\beta)
=\Z_{\J_0}(\Lambda,\beta)
&=
\frac{(4\pi v\Lambda)^{-\lambda}}{2\beta}
\int_{\C_{H_t}} dt \, t^{\lambda-1}e^{-t}
\\
&=
\frac{(4\pi v\Lambda)^{-\lambda}}{2\beta}
\left(1-e^{2\pi i\lambda}\right)\Gamma(\lambda),\label{Eqeq}
\qquad \lambda<0
\\
&=
\frac{\pi i}{\beta}
\frac{e^{i\pi/\beta^2}}
{\Gamma\left(2-1/\beta^2\right)}
(4\pi v\Lambda)^{1-1/\beta^2},
\quad\,\, \beta>1.
\end{align}
where we have used the reflection formula to reach the inverse-Gamma form. Although Eq. \eqref{Eqeq} contains $1-e^{2\pi i\lambda}$, for
$\lambda<0$ this factor belongs to the Gamma-function representation of the integral over $\J_0$. It does not indicate a second contributing thimble.\\

The contrast with the regime $0<\beta<1$ is therefore entirely encoded in the Stokes geometry. For $0<\beta<1$, the Hankel cycle belongs to the relative homology class
\begin{equation}
[\C_H]=[\J_0]-[\J_1],
\end{equation}
and the interference between two adjacent saddle contributions produces the factor
$1-e^{2\pi i/\beta^2}$. For $\beta>1$, the critical points are shifted by half a period in the imaginary direction, and the single Lefschetz thimble $\J_0$ itself connects the two asymptotic sectors of the Hankel contour. The corresponding relative homology class therefore contains only one saddle contribution.

\subsubsection[Case III -- \texorpdfstring{$\beta=1$}{β=1}]
{Case III -- $\boldsymbol{\beta=1}$}

At $\beta=1$, the background charge $q=\beta^{-1}-\beta$ vanishes, and the central charge reduces to $c_L=1$. This particular value constitutes a distinguished point of the timelike-Liouville parametrisation \cite{Strominger:2003fn}. The special character of this point is already visible at the level of the classical zero mode. At the transition between the two regimes discussed above, the classical equation of motion Eq. \eqref{Eq:EOMClassical} becomes the constant-field reduction of the full equation of motion
\begin{equation}
    8 e^{2 \phi} \pi\Lambda =0
\end{equation}
and admits no solution at finite
$\phi$. It can only be satisfied in the limiting sense $\Re\phi \to - \infty$, leading to 
\begin{equation}
    \lim_{\Re\phi\to - \infty} \S_{\rm saddle}[\phi] = 0.
\end{equation}
The constant saddle runs to the boundary of field space, so the usual semi-classical expansion around a finite non-degenerate saddle is not available. In the Picard-Lefschetz picture, this is the boundary degeneration separating the two thimble organisations discussed above: the finite non-degenerate saddle disappears to infinity, and the standard finite-saddle Picard–Lefschetz description ceases to apply. Going to the $t$-plane picture, the equivalent of the saddle being pushed to $\Re \phi\to -\infty$ is that it collides with the origin, which is the point around which the Hankel contour winds, which makes it a genuine degenerate case. At the level of the zero-mode partition function, it reduces to
\begin{equation}
    \Z_{\rm saddle}(\Lambda,1) = \frac12 \int_{\C_{H_t}} \frac{dt}{t} \, e^{-t} = -\pi i,
\end{equation}
for which the branch-cut contribution degenerates into the residue contribution associated with the simple pole at $t=0$. Similarly, taking carefully the $\beta\to 1$ limit of both Case I and II leads to this particular value.\\

From the CFT point of view, the value $c=1$ has also been investigated independently. Runkel and Watts constructed a non-rational interacting $c=1$ conformal field theory as a limit of unitary minimal models \cite{Runkel:2001ng}, and its relation to $c=1$ Liouville theory was subsequently explored in \cite{Fredenhagen:2004cj,Kostov:2005kk}. One should nevertheless be cautious in directly identifying these constructions with timelike Liouville obtained through analytic continuation. In particular, bootstrap analyses of Liouville theory for $c\leq1$ \cite{Ribault:2015sxa} emphasise that the continuation to this regime involves additional subtleties and does not amount to a straightforward definition of timelike Liouville.

\section{Gaussian Multiplicative Chaos Point of View}
\label{Sec:GMC}
Mathematical constructions of timelike Liouville theory, also called \textit{imaginary Liouville theory}, have
provided related contour prescriptions for correlation functions with insertions. Usciati et al. \cite{Usciati:2025cdn} proposed a non-compactified imaginary Liouville theory based on a real Gaussian free field and
argued that it reproduces imaginary DOZZ structure constants without imposing a neutrality constraint. Chatterjee developed a rigorous framework for timelike Liouville using Gaussian
variables with negative variance and proved timelike DOZZ-type formulas under charge-neutrality
assumptions \cite{chatterjee2026rigorousresultstimelikeliouville}. Subsequent work studied exact calculations beyond charge neutrality in special regimes \cite{chatterjee2026exactcalculationschargeneutrality}. More recently, related timelike Liouville observables have also been
studied on surfaces with boundary. In particular, Giribet and Sivilotti computed the disk one-point function using a Coulomb-gas prescription analytically continued in the number of screening operators, obtaining an expression satisfying the expected reflection symmetry, self-duality, and bootstrap shift equations \cite{Giribet:2026gao}.

\subsection{The Dictionary}
We now give an explicit dictionary between our conventions and those used in mathematical constructions of imaginary Liouville field theory (iLFT), whose main ingredient is imaginary Gaussian multiplicative chaos (GMC) \cite{Lacoin:2013joa}. For a surface $\Sigma$ with metric $g$ and scalar curvature $K_g$, the action considered in \cite{Usciati:2025cdn} takes the form
\begin{equation}
    \S_{\rm im. \, L} = \frac{1}{4\pi} \int_\Sigma dv \, \left( |d\tilde\phi|_g^2 + i \tilde q K_g \tilde\phi + \mu e^{i\tilde\beta \tilde\phi}\right) + \frac{1}{2\pi} \int_{\partial\Sigma} dl \, \left( i \tilde q k_g \tilde\phi + \mu_B e^{i\tilde \beta \tilde\phi/2}\right),
    \label{Eq:imLiouvilleAction}
\end{equation}
where $\tilde\phi=\tilde\phi(x)$ is the fundamental field, which is a Gaussian Free Field (GFF), and $\mu,\mu_B\in \mathbb R$ are the bulk and boundary cosmological constants, respectively. The parameter $\tilde q$ is defined in terms of $\tilde \beta\in\mathbb R$ by $\tilde q =\tilde \beta/2- 2/\tilde \beta$, yielding the central charge $c = 1 - 6{\tilde q\,}^2 \leq 1$. Specialising the surface to be the sphere $\Sigma = S^2$ and setting the boundary parameters to zero, their action maps formally to the timelike Liouville action Eq. \eqref{Eq:tLAction} by setting
\begin{align}
&\tilde \beta = 2\beta, & \tilde q(\tilde \beta)  &= -q(\beta), &K_g& =  R, \nonumber\\
&\tilde \phi = -i \phi, & \mu &= 4\pi  \Lambda,
\end{align}
where all the tilded quantities correspond to iLFT and the untilded variables are those used for timelike Liouville in this paper. The idea is then to split the field into a constant zero mode $c$ and a real ultraviolet-regularised GFF $X^\varepsilon_{g_0}$ such that
\begin{equation}
    \tilde \phi(x) = X^\varepsilon_{g_0}(x) + c,
\end{equation}
where $c\in \mathbb C$, separating the constant mode from the fluctuating sector. Since $\partial_ic=0$, the kinetic term depends only on $X_{g_0}^{\varepsilon}$ and is thus independent of the zero mode $c$. Their primary field are vertex operators $V_\alpha$ with scaling dimension $\Delta = \frac \alpha 2 \left(\frac \alpha 2 - \tilde q\right)$, $\alpha\in \mathbb R$, defined as
\begin{equation}
    V_\alpha (x) = \lim_{\varepsilon\to 0} \varepsilon^{-\alpha^2/2} e^{i\alpha \tilde\phi(x)}. 
\end{equation}

\subsection{The Sphere Three-Point Function and Its Zero-Mode Contour}

\begin{figure}
    \centering
    \includegraphics[width=0.8\linewidth]{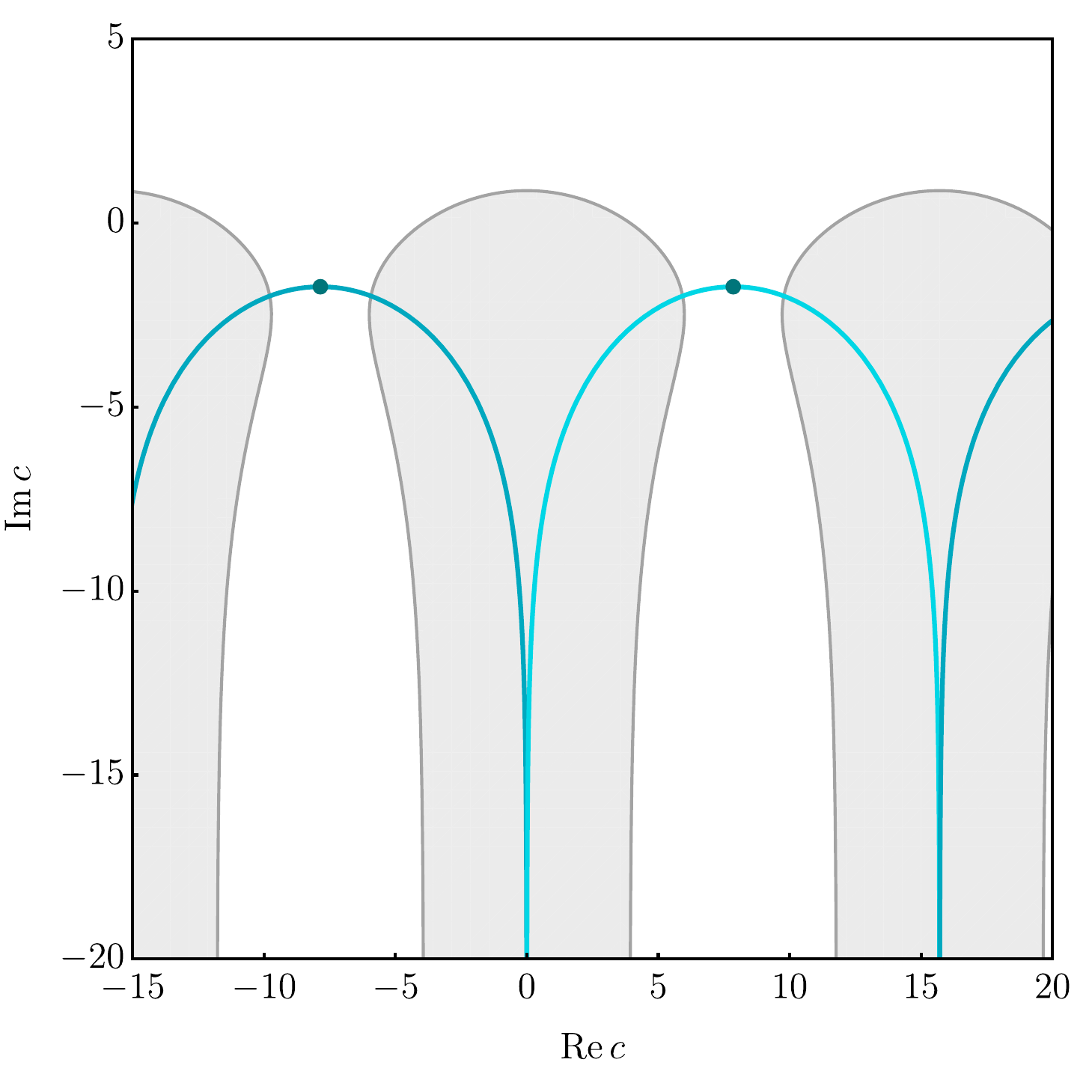}
    \caption{\textbf{Minisuperspace thimble geometry for $\boldsymbol{s>0}$.}
The parameters $\tilde\beta=2/5$, $\mu=1$, and
$\boldsymbol{\alpha}=(-4,-4,-12/5)$ give $s=2$, as in Fig.~2 of
Ref.~\cite{Usciati:2025cdn}. The dark points are the critical points,
the blue curves are the descending thimbles $\mathcal J_n$, and the
shaded regions correspond to $h<0$. The vertical dual thimbles
$\mathcal K_n$ are not shown.}
    \label{Fig:GMCPositiveS}
\end{figure}

\begin{figure}
    \centering
    \includegraphics[width=0.8\linewidth]{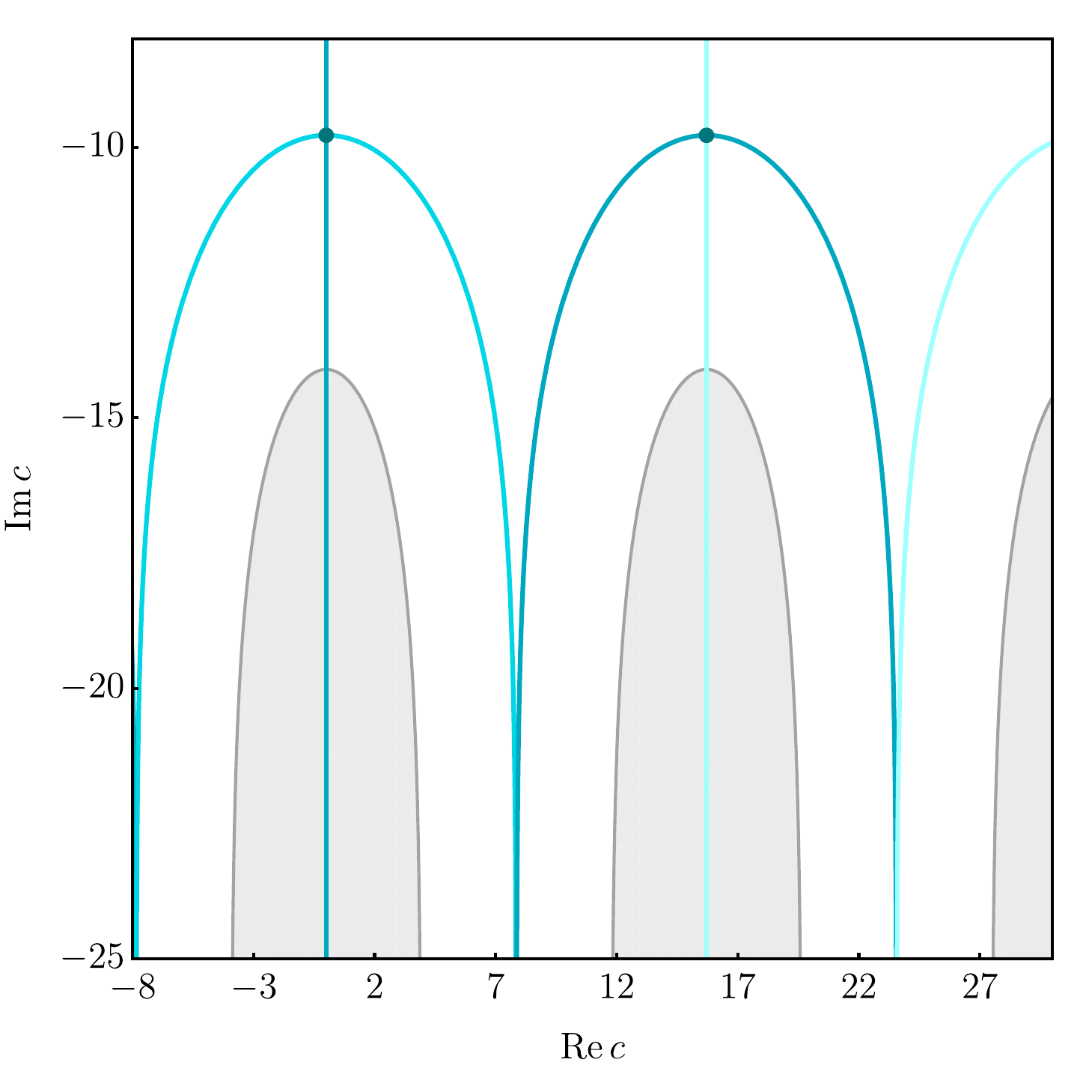}
    \caption{\textbf{Minisuperspace thimble geometry for $\boldsymbol{s<0}$.}
The parameters $\tilde\beta=2/5$, $\mu=1$, and
$\boldsymbol{\alpha}=(4,4,12/5)$ give $s=-50$. The dark points are the
critical points, the vertical blue lines are the descending thimbles
$\mathcal J_n$, and the curved blue branches are their duals
$\mathcal K_n$. The shaded regions correspond to $h<0$.}
 \label{Fig:GMCNegativeS}
\end{figure}

We now specialise to the three-point function on the sphere
\begin{equation}
C_{\boldsymbol{\alpha}}
=
\left\langle
V_{\alpha_1}(0)V_{\alpha_2}(1)V_{\alpha_3}(\infty)
\right\rangle,
\qquad
\boldsymbol{\alpha}=(\alpha_1,\alpha_2,\alpha_3),
\end{equation}
which determines the structure constants of the theory. Denoting
\begin{equation}
\bar{\alpha}=\sum_{j=1}^{3}\alpha_j,
\end{equation}
the probabilistic construction involves the charge-dependent imaginary chaos $M_{\tilde\beta,\boldsymbol{\alpha}}^{S^2}$, whose explicit form is given in Eq.~(13) of Ref.~\cite{Usciati:2025cdn}, together with its Laplace transform
\begin{equation}
\mathcal G_{\tilde\beta,\boldsymbol{\alpha}}^{S^2}(z)
=
\mathbb E\left[
e^{-zM_{\tilde\beta,\boldsymbol{\alpha}}^{S^2}}
\right].
\end{equation}
For the sphere three-point function, the phase-I construction requires
\begin{equation}
    \tilde\beta^2<2,
    \qquad
    \alpha_j>\tilde q,
    \qquad j=1,2,3.
    \label{Eq:SphereGMCBounds}
\end{equation}
These conditions are satisfied by the parameters considered below
\cite{Usciati:2025cdn}. The sphere three-point function can then be written as a zero-mode integral
\begin{equation}
C_{\boldsymbol{\alpha}}
=
\int_{\mathcal U}dc \,
e^{i(\bar{\alpha}-2\tilde q)c}
\mathcal G_{\tilde\beta,\boldsymbol{\alpha}}^{S^2}
\left(\mu e^{i\tilde\beta c}\right),
\label{Eq:SphereThreePointGMC}
\end{equation}
where $\mathcal U$ is the $\mathcal U$-shaped contour
\begin{equation}
-i\infty
\to 0
\to \frac{2\pi}{\tilde\beta}
\to
\frac{2\pi}{\tilde\beta}-i\infty .
\end{equation}
It is convenient to introduce the charge parameter
\begin{equation}
s
=
\frac{2\tilde q-\bar{\alpha}}
{\tilde\beta},
\label{Eq:sSphere}
\end{equation}
so that the exponent governing the zero-mode integral is
\begin{equation}
\mathcal A(c)
=
-i\tilde\beta s c
+
\log
\mathcal G_{\tilde\beta,\boldsymbol{\alpha}}^{S^2}
\left(\mu e^{i\tilde\beta c}\right).
\end{equation}
In the minisuperspace approximation, suppressing the fluctuating GFF
replaces $M_{\tilde\beta,\boldsymbol{\alpha}}^{S^2}$ by a deterministic
constant. Absorbing this constant into $\mu$, the Laplace transform
reduces to
\begin{equation}
\mathcal G_{\tilde\beta,\boldsymbol{\alpha}}^{S^2}
\left(\mu e^{i\tilde\beta c}\right)
\to
e^{-\mu e^{i\tilde\beta c}},
\end{equation}
and therefore
\begin{equation}
\mathcal A_{\mathrm{mini}}(c)
=
-i\tilde\beta s c
-
\mu e^{i\tilde\beta c}.
\label{Eq:MiniExponentGMC}
\end{equation}
Writing $c=x+iy$, the corresponding Morse function is
\begin{equation}
h(x,y)
=
\Re\mathcal A_{\mathrm{mini}}(x+iy)
=
\tilde\beta s y
-
\mu e^{-\tilde\beta y}
\cos(\tilde\beta x).
\label{Eq:MiniMorseGMC}
\end{equation}
Since the integrand is written as $e^{\mathcal A_{\mathrm{mini}}}$, convergent ends must approach regions in which $h\to-\infty$. The shaded regions in Figs.~\ref{Fig:GMCPositiveS} and \ref{Fig:GMCNegativeS} indicate $h<0$. The periodic part of the integrand has period $\frac{2\pi}{\tilde\beta}$, whereas the complete integrand is quasi-periodic:
\begin{equation}
e^{\mathcal A(c+P)}
=
e^{-2\pi i s}e^{\mathcal A(c)}.
\label{Eq:GMCMonodromy}
\end{equation}
The monodromy is therefore trivial whenever $s\in\mathbb Z$, and not only at $s=0$. Nevertheless, $s=0$ remains distinguished: the finite critical points escape to infinity there, and the roles of the curved and vertical flow cycles are exchanged as the sign of $s$ changes. For $s\neq0$, the critical points of the minisuperspace exponent satisfy
\begin{equation}
e^{i\tilde\beta c_n}
=
-\frac{s}{\mu},
\end{equation}
and may be written uniformly as
\begin{equation}
c_n
=
\frac{
2\pi n+\operatorname{arg}(-s)
-i\log(|s|/\mu)}
{\tilde\beta},
\qquad n\in\mathbb Z,
\label{Eq:GMCSaddles}
\end{equation}
where $\operatorname{arg}(-s)=\pi$ for $s>0$ and vanishes for $s<0$. Moreover,
\begin{equation}
\mathcal A_{\mathrm{mini}}''(c_n)
=
-\tilde\beta^2s.
\end{equation}
Consequently, the sign of $s$ determines which constant-phase branch is descending. The two constant-phase cycles through $c_n$ can be parametrised as
\begin{align}
c_n^{\mathrm{curved}}(\theta)
&=
c_n
+
\frac{1}{\tilde\beta}
\left[
\theta
+
i\log\left(\frac{\sin\theta}{\theta}\right)
\right],
\qquad -\pi<\theta<\pi,
\label{Eq:GMCCurvedCycle}\\
c_n^{\mathrm{vertical}}(t)
&=
c_n+\frac{it}{\tilde\beta},
\qquad t\in\mathbb R.
\label{Eq:GMCVerticalCycle}
\end{align}
Their Picard--Lefschetz interpretation is
\begin{equation}
\begin{array}{c|cc}
& \mathcal J_n & \mathcal K_n \\ \hline
s>0 & c_n^{\mathrm{curved}} & c_n^{\mathrm{vertical}}\\
s<0 & c_n^{\mathrm{vertical}} & c_n^{\mathrm{curved}}
\end{array},
\label{Eq:GMCThimbleDictionary}
\end{equation}
where $\mathcal J_n$ denotes the descending thimble and $\mathcal K_n$ its upward-flow dual.

\paragraph{The $\boldsymbol{s>0}$ sector.}
For positive $s$, the saddles lie at
\begin{equation}
c_n
=
\frac{(2n+1)\pi}{\tilde\beta}
-
\frac{i}{\tilde\beta}
\log\left(\frac{s}{\mu}\right).
\end{equation}
The descending thimble is the curved branch in Eq.~\eqref{Eq:GMCCurvedCycle}. In the fundamental strip
\begin{equation}
0\leq\Re c\leq\frac{2\pi}{\tilde\beta},
\end{equation}
the thimble $\mathcal J_0$ joins the two asymptotic ends of the $\mathcal U$-contour, and hence
\begin{equation}
[\mathcal U]=[\mathcal J_0],
\qquad s>0.
\label{Eq:PositiveSDecomposition}
\end{equation}
Figure~\ref{Fig:GMCPositiveS} shows this geometry for the parameters used in Fig.~2 of Ref.~\cite{Usciati:2025cdn},
\begin{equation}
\tilde\beta=\frac25,
\qquad
\mu=1,
\qquad
\boldsymbol{\alpha}
=(-4,-4,-12/5).
\end{equation}
For these values,
\begin{equation}
\tilde q=-\frac{24}{5},
\qquad
s=2,
\qquad
P=5\pi.
\end{equation}
The displayed saddles are located at
\begin{equation}
\Re c=\pm\frac{5\pi}{2},
\qquad
\Im c=-\frac52\log2.
\end{equation}
The figure should be understood as the analytically tractable minisuperspace counterpart of the numerical GMC plot of Ref.~\cite{Usciati:2025cdn}, rather than as a reproduction of the full fluctuating-GFF result.

\paragraph{The $\boldsymbol{s<0}$ sector.}
For negative $s$, the saddles instead lie at
\begin{equation}
c_n
=
\frac{2\pi n}{\tilde\beta}
-
\frac{i}{\tilde\beta}
\log\left(\frac{|s|}{\mu}\right).
\end{equation}
The vertical cycles are now the descending thimbles, whereas the curved branches are the upward dual cycles. Orienting $\mathcal J_n$ toward increasing $\Im c$, the two sides of the $\mathcal U$-contour give
\begin{equation}
[\mathcal U]
=
[\mathcal J_0]-[\mathcal J_1],
\qquad s<0.
\label{Eq:NegativeSDecomposition}
\end{equation}
Figure~\ref{Fig:GMCNegativeS} illustrates this second configuration using
\begin{equation}
\tilde\beta=\frac25,
\qquad
\mu=1,
\qquad
\boldsymbol{\alpha}
=
(4,4,12/5),
\end{equation}
for which
\begin{equation}
s=-50.
\end{equation}
The saddles shown in the figure are located at
\begin{align}
\Re c&=5\pi n, \quad n\in \mathbb Z,\\
\Im c&=-\frac52\log50.
\end{align}
The plot is a zoom around these saddles. The vertical descending thimbles continue outside the displayed window toward both $\Im c\to-\infty$ and $\Im c\to+\infty$.

\subsection{GMC, Stokes, and Charge Neutrality}

It is important to distinguish three conditions that enter the preceding
discussion. First, the phase-I imaginary-GMC construction requires
\begin{equation}
    \tilde\beta^2<2
    \qquad\Leftrightarrow\qquad
    \beta^2<\frac12.
\end{equation}
This condition concerns the existence of the non-zero-mode random
distribution and is independent of the choice of zero-mode integration
cycle. At and beyond $\tilde\beta^2=2$, this phase-I normalisation no longer produces the imaginary-chaos distribution entering
Eq.~\eqref{Eq:SphereThreePointGMC}. Alternative critical and phase-III
normalisations can lead to white-noise-type limits
\cite{lacoin2020convergencelawcomplexgaussian}, but these define
different random objects. Consequently, the direct comparison with the probabilistic construction used above is restricted to
$\tilde\beta^2<2$, whereas the zero-mode Picard--Lefschetz analysis can
be continued formally beyond this range.\\

Second, the Picard--Lefschetz decomposition of the zero-mode contour is controlled by
\begin{equation}
    s
    =
    \frac{2\tilde q-\sum_j\alpha_j}{\tilde\beta},
\end{equation}
rather than by $\tilde\beta$ alone. For the zero-point function,
$\sum_j\alpha_j=0$, and the dictionary
$\tilde\beta=2\beta$ and $\tilde q=-q$ gives
\begin{equation}
    s_0
    =
    1-\frac{1}{\beta^2}.
\end{equation}
The corresponding change in thimble decomposition occurs at
$s_0=0$, or equivalently at $\beta^2=1$. For correlation functions, the insertion charges shift $s$, so either sign of $s$ can
occur within the phase-I GMC range.\\

Finally, charge neutrality is a separate discrete condition. One has
\begin{equation}
    s
    =
    \frac{2\tilde q-\sum_j\alpha_j}{\tilde\beta}
    \equiv w_n.
\end{equation}
Thus the finite-screening sector is characterised by
$s=w_n\in\mathbb Z_{\geq0}$. The rigorous charge-neutral result of
Ref.~\cite{chatterjee2026rigorousresultstimelikeliouville} assumes
$w_n\in\mathbb Z_{>0}$ and $n\geq3$. This discrete condition should not be identified with either the GMC threshold
$\tilde\beta^2=2$ or the zero-point Stokes transition $s_0=0$. The distinction is particularly transparent for the zero-point function. Formally setting $n=0$ gives
\begin{equation}
    w_0=s_0
    =
    1-\frac{1}{\beta^2}.
\end{equation}
For $0<\beta<1$, this quantity is negative, while for $\beta>1$ it
satisfies $0<w_0<1$. At $\beta=1$, one has $w_0=0$. Thus, for every
finite $\beta>0$, the sphere partition function lies outside the
positive-integer sector assumed in the rigorous charge-neutral result. The transition at $s_0=0$ is a change in the zero-mode thimble decomposition, not an instance of the charge-neutral theorem. The special coupling $b=1/\sqrt{2}$ considered in
Ref.~\cite{chatterjee2026exactcalculationschargeneutrality} provides a
further useful comparison. Under $\tilde\beta=2b$, it corresponds to
$\tilde\beta=\sqrt{2}$, precisely the boundary of the phase-I GMC
regime, whereas $w_0=-1$. This coincidence of coupling values should
not be interpreted as an identification of the GMC threshold with charge neutrality.

\section{Conclusion and Discussion}
\label{Sec:Conclusion}

This work was devoted to understanding the saddle content of the two-sphere partition function in timelike Liouville theory, solving the apparent mismatch between two complementary descriptions of the partition function. On the one hand, the gravitational path integral evaluated around the real saddle reproduces the universal small-$\beta$ expansion of the analytically continued DOZZ expression. On the other hand, the latter contains the additional factor
\begin{equation}
1-e^{2\pi i/\beta^2},
\end{equation}
naturally suggesting the contribution of a second saddle. Our goal was to determine whether this additional contribution could be recovered directly from the integration cycle of the timelike-Liouville zero mode.\\

To address this question, we performed a Picard--Lefschetz analysis of the zero-mode integral, taking as input the Hankel-type contour motivated by the inverse-Gamma structure of the analytically continued integral and by related contour prescriptions for charged correlators in imaginary Liouville theory. The choice of contour does not follow uniquely from convergence, nor does the zero-mode calculation establish the complete normalisation of the full partition function. The zero-mode problem exhibits two non-degenerate regimes,  separated by a boundary degeneration at $\beta=1$. At this point, the background charge vanishes and the finite non-degenerate saddle is pushed to the boundary of field space, $\Re\phi\to-\infty$. As $\beta$ crosses this critical value, the Lefschetz-thimble decomposition of the same integration cycle reorganises.\\

For $0<\beta<1$, the critical points form an infinite family related by imaginary shifts of the Liouville zero mode. In this regime, the Hankel contour is not homologous to the thimble associated with the real saddle alone. Rather, its relative-homology decomposition is
\begin{equation}
[\C_H]=[\J_0]-[\J_1].
\end{equation}
The non-perturbative factor appearing in the analytically-continued DOZZ expression therefore admits a direct Picard--Lefschetz interpretation: it reflects the relative homology class of the Hankel-type contour. The second saddle does not need to be introduced by hand in order to reproduce the DOZZ result, but follows from the decomposition of the chosen integration cycle.\\

For $\beta>1$, the situation changes drastically. The same integration cycle is instead represented by a single relevant thimble,
\begin{equation}
[\C_H]=[\J_0].
\end{equation}
This reorganisation is conditional on the chosen integration cycle. For $0<\beta<1$, the real zero-mode contour also converges,
whereas for $\beta>1$ its endpoint divergence requires an additional prescription without selecting a unique cycle. Consequently, our analysis explains the saddle factor generated by the Hankel-type cycle without claiming a unique physical definition of the full timelike partition function.\\

The probabilistic imaginary-Liouville construction provides a useful comparison with our contour choice, but its results for charged
correlators do not determine the sphere zero-point prescription. In the corresponding conventions, the monodromy of the zero-mode integrand is controlled by the charge parameter $s$. In the formal zero-point case,
\begin{equation}
s_0=1-\frac{1}{\beta^2},
\end{equation}
whose sign changes precisely at $\beta=1$. The formal zero-point specialisation of the probabilistic monodromy parameter singles out the same value. This extrapolates beyond the phase-I iGMC range.\\

Several extensions of this analysis would be worthwhile. A natural first step would be to include vertex-operator insertions. The monodromy parameter would then depend explicitly on the insertion charges, and the interplay between the Stokes decomposition, probabilistic contour prescriptions and charge-neutral sectors should become considerably richer. The two-sphere partition function considered here provides the simplest setting in which these ingredients can be disentangled.\\

Another question concerns the formal $\beta\leftrightarrow1/\beta$ duality. At the level of the background charge
\begin{equation}
q=\frac{1}{\beta}-\beta,
\end{equation}
this transformation exchanges $q\leftrightarrow-q$ and relates the regimes $0<\beta<1$ and $\beta>1$. Our analysis, however, shows that the saddle configurations and Stokes geometries on the two sides of $\beta=1$ are qualitatively different. It is therefore not immediate that the duality extends to the full Lefschetz-thimble decomposition of the path integral. It would be interesting to determine whether it induces a non-trivial map between admissible integration cycles, thimbles and intersection coefficients. Such a correspondence could clarify whether the $\beta>1$ regime should be regarded as a genuinely distinct non-perturbative sector or as a dual description of the conventional $0<\beta<1$ branch.\\

The zero-mode integral also provides a simple setting in which the relation between the saddle structure and resurgence can be made more concrete. Going to the $t$-plane for $0<\beta<1$, or $\lambda>0$, we obtained Eq. \eqref{Eq:GammaRepBetasmall}. In the semiclassical limit $\beta\to0$, one has $\lambda\to+\infty$, and the saddle-point expansion is governed by the Stirling series
\begin{equation}
\log\Gamma(\lambda)
\sim
\left(\lambda-\frac{1}{2}\right)\log\lambda-\lambda
+\frac{1}{2}\log(2\pi)
+\sum_{n=1}^{\infty}
\frac{B_{2n}}{2n(2n-1)\lambda^{2n-1}}.
\end{equation}
Since
\begin{equation}
B_{2n}
\sim
(-1)^{n-1}\frac{2(2n)!}{(2\pi)^{2n}},
\end{equation}
this semiclassical expansion is Gevrey-one type. Hence, the Borel transform of
\begin{equation}
    \phi(\lambda) = \sum_{n=1}^{\infty}
\frac{B_{2n}}{2n(2n-1)}\lambda^{-2n+1}
\end{equation}
is 
\begin{equation}
\hat{\phi}(\zeta)
=
\frac{1}{\zeta^2}
\left(
\frac{\zeta}{2}\coth\frac{\zeta}{2}-1
\right),
\end{equation}
and has singularities at
\begin{equation}
\zeta_n=2\pi i n,
\qquad
n\in\mathbb Z\setminus{\{0\}}.
\end{equation}
These singularities have a direct counterpart in the saddle structure of the original zero-mode integral. The critical points satisfy
\begin{equation}
\varphi_n=\varphi_0+\frac{i\pi n}{\beta},
\end{equation}
while their action differences are
\begin{equation}
S(\varphi_n)-S(\varphi_0)
=
-2\pi i n\lambda = A_n \lambda,
\end{equation}
In terms of $\beta$, the relative saddle contributions take the characteristic form
\begin{equation}
e^{-A_n\lambda }=e^{A_n}e^{-A_n/\beta^2}\equiv e^{-A_n/\beta^2},
\qquad
A_n=\pm2\pi i n.
\end{equation}
Thus, up to conventions, the locations of the Borel singularities coincide with the coefficients $A_n$ governing the action differences between the different logarithmic lifts of the saddle. At the level of the zero-mode sector, this provides a concrete resurgent relation: the large-order behaviour of the perturbative expansion around one saddle retains information about the other saddles.\\

This observation should be distinguished from a complete transseries description of timelike Liouville theory. In particular, for $\beta>1$, the large-$\lambda$ Stirling expansion no longer defines the relevant semiclassical regime. The exact Hankel representation nevertheless continues to retain the global information associated with the contour and its different logarithmic sheets. It would therefore be interesting to understand whether the rearrangement of thimbles across $\beta=1$ can be embedded into a broader resurgent description.\\

More broadly, this perspective may be relevant to the longstanding problem of analytically continuing spacelike Liouville theory to its timelike counterpart. A naive continuation of the parameters, $b\to i\beta$, does not by itself specify the appropriate integration cycle of the timelike theory. Indeed, reproducing timelike amplitudes from the original Liouville path integral is known to require a non-trivial modification of the integration contour \cite{Harlow_2011}. From the viewpoint developed here, analytic continuation in parameter space may cross Stokes walls and thereby change the decomposition of the integration cycle into Lefschetz thimbles. Continuing a single saddle contribution would then fail to capture the accompanying change in intersection coefficients and, more generally, the global information carried by the contour. The zero-mode model thus provides an explicit example in which the analytic continuation of a gravitational path integral is inseparable from the global topology of its integration cycle.

\section*{Acknowledgements}
I would particularly like to thank Beatrix Mühlmann for introducing me to the de Sitter world, for suggesting this problem, as well as for the time that she devoted to me for ansewering my questions and for insightful discussions. I am grateful to Gaston Giribet and Dionysios Anninos for their helpful comments and suggestions on the draft of this paper. I also thank Andrés Collinucci and Michele Lenzi for related discussions, and in particular Stathis Vitouladitis for proofreading this paper. I thank the 31st WE-Heraeus Summer School ``Saalburg" for creating a stimulating environment and in particular Beatrix Mühlmann for her lectures and discussions on de Sitter and related topics. Finally, I thank the organisers and participants of the ``de Sitter day" at the University of Mons for the stimulating workshop and discussions. ED is a Research Fellow of the Fonds de la Recherche Scientifique F.R.S.-FNRS (Belgium). The author is a member of BLU-ULB, the interfaculty research group focusing on space research at ULB. This work is supported by the F.R.S.-FNRS (Belgium) through convention IISN 4.4514.08 and benefited from the support of the Solvay Family.

\bibliographystyle{JHEP}
\bibliography{biblio}

\providecommand{\href}[2]{#2}\begingroup\raggedright\begin{thebibliography}{10}

\bibitem{PolyakovNonCriticalStrings}
A.~M. Polyakov, \emph{{Quantum Geometry of Bosonic Strings}}, \href{http://dx.doi.org/10.1016/0370-2693(81)90743-7}{\emph{Phys. Lett. B} {\bfseries 103} (1981) 207--210}.

\bibitem{Distler:1988jt}
J.~Distler and H.~Kawai, \emph{{Conformal Field Theory and 2D Quantum Gravity}}, \href{http://dx.doi.org/10.1016/0550-3213(89)90354-4}{\emph{Nucl. Phys. B} {\bfseries 321} (1989) 509--527}.

\bibitem{David:1988hj}
F.~David, \emph{{Conformal Field Theories Coupled to 2D Gravity in the Conformal Gauge}}, \href{http://dx.doi.org/10.1142/S0217732388001975}{\emph{Mod. Phys. Lett. A} {\bfseries 3} (1988) 1651}.

\bibitem{Moore:1991zv}
G.~W. Moore, M.~R. Plesser and S.~Ramgoolam, \emph{{Exact S matrix for 2-D string theory}}, \href{http://dx.doi.org/10.1016/0550-3213(92)90020-C}{\emph{Nucl. Phys. B} {\bfseries 377} (1992) 143--190}, [\href{https://arxiv.org/abs/hep-th/9111035}{{\ttfamily hep-th/9111035}}].

\bibitem{Seiberg:2004at}
N.~Seiberg and D.~Shih, \emph{{Minimal string theory}}, \href{http://dx.doi.org/10.1016/j.crhy.2004.12.007}{\emph{Comptes Rendus Physique} {\bfseries 6} (2005) 165--174}, [\href{https://arxiv.org/abs/hep-th/0409306}{{\ttfamily hep-th/0409306}}].

\bibitem{Collier:2023cyw}
S.~Collier, L.~Eberhardt, B.~M{\"u}hlmann and V.~A. Rodriguez, \emph{{The Virasoro minimal string}}, \href{http://dx.doi.org/10.21468/SciPostPhys.16.2.057}{\emph{SciPost Phys.} {\bfseries 16} (2024) 057}, [\href{https://arxiv.org/abs/2309.10846}{{\ttfamily 2309.10846}}].

\bibitem{Collier:2024kmo}
S.~Collier, L.~Eberhardt, B.~M{\"u}hlmann and V.~A. Rodriguez, \emph{{Complex Liouville String}}, \href{http://dx.doi.org/10.1103/k74n-s63l}{\emph{Phys. Rev. Lett.} {\bfseries 134} (2025) 251602}, [\href{https://arxiv.org/abs/2409.17246}{{\ttfamily 2409.17246}}].

\bibitem{Knizhnik:1988ak}
V.~G. Knizhnik, A.~M. Polyakov and A.~B. Zamolodchikov, \emph{{Fractal Structure of 2D Quantum Gravity}}, \href{http://dx.doi.org/10.1142/S0217732388000982}{\emph{Mod. Phys. Lett. A} {\bfseries 3} (1988) 819}.

\bibitem{Seiberg:1990eb}
N.~Seiberg, \emph{{Notes on quantum Liouville theory and quantum gravity}}, \href{http://dx.doi.org/10.1143/PTPS.102.319}{\emph{Prog. Theor. Phys. Suppl.} {\bfseries 102} (1990) 319--349}.

\bibitem{zamolodchikov_structure_1996}
A.~B. Zamolodchikov and A.~B. Zamolodchikov, \emph{Structure {Constants} and {Conformal} {Bootstrap} in {Liouville} {Field} {Theory}}, \href{http://dx.doi.org/10.1016/0550-3213(96)00351-3}{\emph{Nuclear Physics B} {\bfseries 477} (Oct., 1996) 577--605}.

\bibitem{Dorn_1994}
H.~Dorn and H.-J. Otto, \emph{Two- and three-point functions in liouville theory}, \href{http://dx.doi.org/10.1016/0550-3213(94)00352-1}{\emph{Nuclear Physics B} {\bfseries 429} (Oct., 1994) 375–388}.

\bibitem{Teschner:2001rv}
J.~Teschner, \emph{{Liouville theory revisited}}, \href{http://dx.doi.org/10.1088/0264-9381/18/23/201}{\emph{Class. Quant. Grav.} {\bfseries 18} (2001) R153--R222}, [\href{https://arxiv.org/abs/hep-th/0104158}{{\ttfamily hep-th/0104158}}].

\bibitem{Harlow_2011}
D.~Harlow, J.~Maltz and E.~Witten, \emph{{Analytic Continuation of Liouville Theory}}, \href{http://dx.doi.org/10.1007/JHEP12(2011)071}{\emph{JHEP} {\bfseries 12} (2011) 071}, [\href{https://arxiv.org/abs/1108.4417}{{\ttfamily 1108.4417}}].

\bibitem{Giribet:2011zx}
G.~Giribet, \emph{{On the timelike Liouville three-point function}}, \href{http://dx.doi.org/10.1103/PhysRevD.85.086009}{\emph{Phys. Rev. D} {\bfseries 85} (2012) 086009}, [\href{https://arxiv.org/abs/1110.6118}{{\ttfamily 1110.6118}}].

\bibitem{Ribault:2015sxa}
S.~Ribault and R.~Santachiara, \emph{{Liouville theory with a central charge less than one}}, \href{http://dx.doi.org/10.1007/JHEP08(2015)109}{\emph{JHEP} {\bfseries 08} (2015) 109}, [\href{https://arxiv.org/abs/1503.02067}{{\ttfamily 1503.02067}}].

\bibitem{Mertens:2020hbs}
T.~G. Mertens and G.~J. Turiaci, \emph{{Liouville quantum gravity -- holography, JT and matrices}}, \href{http://dx.doi.org/10.1007/JHEP01(2021)073}{\emph{JHEP} {\bfseries 01} (2021) 073}, [\href{https://arxiv.org/abs/2006.07072}{{\ttfamily 2006.07072}}].

\bibitem{anninos_two-sphere_2021}
D.~Anninos, T.~Bautista and B.~Mühlmann, \emph{The two-sphere partition function in two-dimensional quantum gravity}, \href{http://dx.doi.org/10.1007/JHEP09(2021)116}{\emph{Journal of High Energy Physics} {\bfseries 2021} (Sept., 2021) 116}.

\bibitem{anninos_remarks_2025}
D.~Anninos, C.~Baracco and B.~M{\"u}hlmann, \emph{{Remarks on 2D quantum cosmology}}, \href{http://dx.doi.org/10.1088/1475-7516/2024/10/031}{\emph{JCAP} {\bfseries 10} (2024) 031}, [\href{https://arxiv.org/abs/2406.15271}{{\ttfamily 2406.15271}}].

\bibitem{Usciati:2025cdn}
R.~Usciati, C.~Guillarmou, R.~Rhodes and R.~Santachiara, \emph{{Probabilistic Construction of Noncompactified Imaginary Liouville Field Theory}}, \href{http://dx.doi.org/10.1103/d723-gk4m}{\emph{Phys. Rev. Lett.} {\bfseries 136} (2026) 031601}, [\href{https://arxiv.org/abs/2505.09390}{{\ttfamily 2505.09390}}].

\bibitem{Anninos_2026}
D.~Anninos, T.~Hertog and J.~Karlsson, \emph{{Quantum Liouville cosmology}}, \href{http://dx.doi.org/10.1088/1475-7516/2026/06/065}{\emph{JCAP} {\bfseries 06} (2026) 065}, [\href{https://arxiv.org/abs/2512.15969}{{\ttfamily 2512.15969}}].

\bibitem{chatterjee2026rigorousresultstimelikeliouville}
S.~Chatterjee, \emph{{Rigorous results for timelike liouville field theory}}, \href{http://dx.doi.org/10.1017/fms.2026.10223}{\emph{Forum Math. Sigma} {\bfseries 14} (2026) e71}, [\href{https://arxiv.org/abs/2504.02348}{{\ttfamily 2504.02348}}].

\bibitem{chatterjee2026exactcalculationschargeneutrality}
S.~Chatterjee, \emph{{Exact calculations beyond charge neutrality in timelike Liouville field theory}},  \href{https://arxiv.org/abs/[2601.19097]}{{\ttfamily [2601.19097]}}.

\bibitem{Giribet:2026gao}
G.~Giribet and B.~Sivilotti, \emph{{The disk 1-point function in timelike Liouville theory}}, \href{http://dx.doi.org/10.1007/JHEP07(2026)158}{\emph{JHEP} {\bfseries 07} (2026) 158}, [\href{https://arxiv.org/abs/2603.12084}{{\ttfamily 2603.12084}}].

\bibitem{Cercle:2026ihl}
B.~Cercl{\'e} and R.~Usciati, \emph{{On the local conformal structure of Imaginary Liouville theory}},  \href{https://arxiv.org/abs/2608.02543}{{\ttfamily 2608.02543}}.

\bibitem{CarneirodaCunha:2003mxy}
B.~Carneiro~da Cunha and E.~J. Martinec, \emph{{Closed string tachyon condensation and world sheet inflation}}, \href{http://dx.doi.org/10.1103/PhysRevD.68.063502}{\emph{Phys. Rev. D} {\bfseries 68} (2003) 063502}, [\href{https://arxiv.org/abs/hep-th/0303087}{{\ttfamily hep-th/0303087}}].

\bibitem{Bautista:2019jau}
T.~Bautista, A.~Dabholkar and H.~Erbin, \emph{{Quantum Gravity from Timelike Liouville theory}}, \href{http://dx.doi.org/10.1007/JHEP10(2019)284}{\emph{JHEP} {\bfseries 10} (2019) 284}, [\href{https://arxiv.org/abs/1905.12689}{{\ttfamily 1905.12689}}].

\bibitem{muhlmann_two-sphere_2021}
B.~Mühlmann, \emph{The two-sphere partition function in two-dimensional quantum gravity at fixed area}, \href{http://dx.doi.org/10.1007/JHEP09(2021)189}{\emph{Journal of High Energy Physics} {\bfseries 2021} (Sept., 2021) 189}.

\bibitem{Giribet:2022cvw}
G.~Giribet and M.~Leoni, \emph{{2D quantum gravity partition function on the fluctuating sphere}}, \href{http://dx.doi.org/10.1007/JHEP09(2022)126}{\emph{JHEP} {\bfseries 09} (2022) 126}, [\href{https://arxiv.org/abs/2206.05546}{{\ttfamily 2206.05546}}].

\bibitem{Gibbons:1978ac}
G.~W. Gibbons, S.~W. Hawking and M.~J. Perry, \emph{{Path Integrals and the Indefiniteness of the Gravitational Action}}, \href{http://dx.doi.org/10.1016/0550-3213(78)90161-X}{\emph{Nucl. Phys. B} {\bfseries 138} (1978) 141--150}.

\bibitem{polchinski_phase_1989}
J.~Polchinski, \emph{The phase of the sum over spheres}, \href{http://dx.doi.org/10.1016/0370-2693(89)90387-0}{\emph{Physics Letters B} {\bfseries 219} (Mar., 1989) 251--257}.

\bibitem{Maldacena:2024spf}
J.~Maldacena, \emph{{Real observers solving imaginary problems}},  \href{https://arxiv.org/abs/2412.14014}{{\ttfamily 2412.14014}}.

\bibitem{Zamolodchikov:2005fy}
A.~B. Zamolodchikov, \emph{{Three-point function in the minimal Liouville gravity}}, \href{http://dx.doi.org/10.1007/s11232-005-0003-3}{\emph{Theor. Math. Phys.} {\bfseries 142} (2005) 183--196}, [\href{https://arxiv.org/abs/hep-th/0505063}{{\ttfamily hep-th/0505063}}].

\bibitem{Kostov:2005av}
I.~K. Kostov and V.~B. Petkova, \emph{{Non-rational 2-D quantum gravity. I. World sheet CFT}}, \href{http://dx.doi.org/10.1016/j.nuclphysb.2007.02.014}{\emph{Nucl. Phys. B} {\bfseries 770} (2007) 273--331}, [\href{https://arxiv.org/abs/hep-th/0512346}{{\ttfamily hep-th/0512346}}].

\bibitem{Ribault:2023vqs}
S.~Ribault and I.~Tsiares, \emph{{On the Virasoro fusion kernel at $c=25$}}, \href{http://dx.doi.org/10.21468/SciPostPhys.17.2.058}{\emph{SciPost Phys.} {\bfseries 17} (2024) 058}, [\href{https://arxiv.org/abs/2310.09334}{{\ttfamily 2310.09334}}].

\bibitem{muhlmann_two-sphere_2022}
B.~Mühlmann, \emph{The two-sphere partition function from timelike {Liouville} theory at three-loop order}, \href{http://dx.doi.org/10.1007/JHEP05(2022)057}{\emph{Journal of High Energy Physics} {\bfseries 2022} (May, 2022) 57}.

\bibitem{Lacoin:2013joa}
H.~Lacoin, R.~Rhodes and V.~Vargas, \emph{{Complex Gaussian multiplicative chaos}}, \href{http://dx.doi.org/10.1007/s00220-015-2362-4}{\emph{Commun. Math. Phys.} {\bfseries 337} (2015) 569--632}, [\href{https://arxiv.org/abs/1307.6117}{{\ttfamily 1307.6117}}].

\bibitem{guillarmou2023compactifiedimaginaryliouvilletheory}
C.~Guillarmou, A.~Kupiainen and R.~Rhodes, \emph{{Compactified imaginary Liouville theory}}, \href{http://dx.doi.org/10.1090/cams/54}{\emph{Commun. Am. Math. Soc.} {\bfseries 5} (2025) 571--694}, [\href{https://arxiv.org/abs/2310.18226}{{\ttfamily 2310.18226}}].

\bibitem{Cao_2023}
X.~Cao, R.~Santachiara and R.~Usciati, \emph{{On the analytical continuation of lattice Liouville theory}}, \href{http://dx.doi.org/10.1007/JHEP03(2023)061}{\emph{JHEP} {\bfseries 03} (2023) 061}, [\href{https://arxiv.org/abs/2301.07454}{{\ttfamily 2301.07454}}].

\bibitem{Polchinski:1989fn}
J.~Polchinski, \emph{{A Two-Dimensional Model for Quantum Gravity}}, \href{http://dx.doi.org/10.1016/0550-3213(89)90184-3}{\emph{Nucl. Phys. B} {\bfseries 324} (1989) 123--140}.

\bibitem{Pasquetti_2010}
S.~Pasquetti and R.~Schiappa, \emph{Borel and stokes nonperturbative phenomena in topological string theory and c=1 matrix models}, \href{http://dx.doi.org/10.1007/s00023-010-0044-5}{\emph{Annales Henri Poincaré} {\bfseries 11} (June, 2010) 351–431}.

\bibitem{witten2010analyticcontinuationchernsimonstheory}
E.~Witten, \emph{{Analytic Continuation Of Chern-Simons Theory}}, {\emph{AMS/IP Stud. Adv. Math.} {\bfseries 50} (2011) 347--446}, [\href{https://arxiv.org/abs/1001.2933}{{\ttfamily 1001.2933}}].

\bibitem{Gutperle_2003}
M.~Gutperle and A.~Strominger, \emph{Timelike boundary liouville theory}, \href{http://dx.doi.org/10.1103/physrevd.67.126002}{\emph{Physical Review D} {\bfseries 67} (June, 2003) }.

\bibitem{Strominger:2003fn}
A.~Strominger and T.~Takayanagi, \emph{{Correlators in time - like bulk Liouville theory}}, \href{http://dx.doi.org/10.4310/ATMP.2003.v7.n2.a6}{\emph{Adv. Theor. Math. Phys.} {\bfseries 7} (2003) 369--379}, [\href{https://arxiv.org/abs/hep-th/0303221}{{\ttfamily hep-th/0303221}}].

\bibitem{Runkel:2001ng}
I.~Runkel and G.~M.~T. Watts, \emph{{A Nonrational CFT with c = 1 as a limit of minimal models}}, \href{http://dx.doi.org/10.1088/1126-6708/2001/09/006}{\emph{JHEP} {\bfseries 09} (2001) 006}, [\href{https://arxiv.org/abs/hep-th/0107118}{{\ttfamily hep-th/0107118}}].

\bibitem{Fredenhagen:2004cj}
S.~Fredenhagen and V.~Schomerus, \emph{{Boundary Liouville theory at c = 1}}, \href{http://dx.doi.org/10.1088/1126-6708/2005/05/025}{\emph{JHEP} {\bfseries 05} (2005) 025}, [\href{https://arxiv.org/abs/hep-th/0409256}{{\ttfamily hep-th/0409256}}].

\bibitem{Kostov:2005kk}
I.~K. Kostov and V.~B. Petkova, \emph{{Bulk correlation functions in 2-D quantum gravity}}, \href{http://dx.doi.org/10.1007/s11232-006-0011-y}{\emph{Theor. Math. Phys.} {\bfseries 146} (2006) 108--118}, [\href{https://arxiv.org/abs/hep-th/0505078}{{\ttfamily hep-th/0505078}}].

\bibitem{lacoin2020convergencelawcomplexgaussian}
H.~Lacoin, \emph{Convergence in law for complex gaussian multiplicative chaos in phase iii},  2020.

\end{thebibliography}\endgroup
\end{document}